\documentclass[aps,pra, superscriptaddress,reprint]{revtex4-1}

\usepackage{placeins}
\usepackage{adjustbox}
\usepackage{graphicx}
\usepackage{booktabs, makecell}
\usepackage{lipsum}
\usepackage{physics}
\usepackage[english]{babel}
\usepackage{subcaption}
\usepackage{tabularx}
\usepackage{hyperref}
\hypersetup{
    colorlinks=true,
    linkcolor=blue,
    filecolor=magenta,      
    urlcolor=cyan,
}
 
\usepackage{dcolumn}
\usepackage{listings}
\usepackage{floatrow}
\newfloatcommand{capbtabbox}{table}[][\FBwidth]

\usepackage{blindtext}
\usepackage{graphicx}  
\usepackage{dcolumn}   
\usepackage{bm}        
\usepackage{amsmath,amssymb}
\usepackage{color}
\usepackage[dvipsnames]{xcolor}
\definecolor{purple}{rgb}{0.58,0.0,0.83}
\usepackage{caption}
\usepackage[toc,page]{appendix}
\usepackage[utf8]{inputenc}
\usepackage{tabularx}
\usepackage{schemata}
\usepackage{subfiles}
\AtBeginDocument{\selectlanguage{english}}

\begin{document}
\title{Simulating open quantum systems using noise models and NISQ devices with error mitigation}
\author{Mainak Roy}
\author{Jessica John Britto}
\affiliation{Indian Institute of Technology Kharagpur, Kolkata, India}
\author{Ryan Hill}
\affiliation{qBraid, 60615, Chicago, United States}
\author{Victor Onofre}
\email{vonofre68@gmail.com}
\affiliation{Quantum Open Source Foundation}

\date{\today}

\begin{abstract}

\subsection*{}
\noindent
 In this work, we present simulations of two Open Quantum System models, Collisional and Markovian Reservoir, with noise simulations, the IBM devices (\texttt{ibmq\_kyoto}, \texttt{ibmq\_osaka}) and with the OQC device Lucy. Extending the results of \textit{García-Pérez, et al. [npj Quantum Information 6.1 (2020): 1]}. Using the Mitiq toolkit, we apply Zero-Noise extrapolation (ZNE), an error mitigation technique and analyze their deviation from the theoretical results for the models under study. For both models, by applying ZNE, we were able to reduce the error and overlap it with the theoretical results. All our simulations and experiments were done in the \texttt{qBraid} environment.

\textbf{\textit{Keywords}}---Open Quantum Systems, NISQ devices, Quantum Error Mitigation
\end{abstract}
 
\maketitle


\section{Introduction}

In the field of quantum computing, all real quantum systems are open and can't be perfectly isolated. As a result, there is always an external environment affecting the system, and the extent of this impact is influenced by various factors that are difficult to control. To understand the behavior of these open quantum systems and control it, accurate simulation is crucial, especially in the gate-based approach. 

In this work, we simulated the correlated collisional model and Markovian reservoir engineering (MRE) based on the results presented in \cite{garcia2020ibm}. The models were simulated using Oxford Quantum Computers (OQC)'s  and the latest IBM quantum systems (\texttt{ibmq\_kyoto} and \texttt{ibmq\_osaka}, IBM Quantum Eagle type processors), both based on superconducting qubits . To update and extend the results from the original paper\cite{garcia2020ibm} using IBM Hardware 2020. 

We also did an analysis of the effect of noise on the quantum circuits with depolarising noise models. In the majority of the cases that we considered, the effects of noise were overcome with Zero Noise Extrapolation (ZNE) \cite{giurgica2020digital} using the Mitiq toolkit \cite{larose2022mitiq}.  We run all the experiments in the \texttt{qBraid-SDK} environment (a Python toolkit for cross-framework abstraction, transpilation, and execution of quantum programs on hardware and simulators \cite{Hill_qBraid-SDK_Python_toolkit_2023}). The circuits were build with Cirq \cite{cirq_developers_2023_10247207}, using Qiskit \cite{Qiskit} to run on the IBM devices with the IBM Quantum Services \cite{IBM_quantum}, and AWS braket \cite{braket} for OQC Lucy \cite{OQC_quantum}. 

\section{Open quantum systems circuit models} \label{Sect:2} 

\subsection{Markovian reservoir engineering}

This model was developed by \cite{garcia2020ibm}, with the initial ideas for ion traps presented in \cite{barreiro2011open}. A carefully designed environment can give rise to interactions that drive the system to prepare a maximally entangled state as a result of the dissipative open system. The four Bell states can be identified as the eigenstates of the operators $\sigma_x^{(1)} \otimes \sigma_x^{(2)}$ and $\sigma_z^{(1)} \otimes \sigma_z^{(2)}$ as shown in table \ref{table:1}.
We can design two channels, the XX pump and the ZZ pump. The action of each is to pump states from the +1 eigenspace to the $-1$ eigenspace of the corresponding operator. The only state belonging to the $-1$ eigenspace of both $\sigma_x^{(1)} \otimes \sigma_x^{(2)}$ and $\sigma_z^{(1)} \otimes \sigma_z^{(2)}$ is $\ket{\psi^{-}}$. If we compose the two channels and apply them together, we can pump everything to a single state, which is $\ket{\psi^{-}}$.

\begin{table}[h!]
\centering
\begin{tabular}{ | m{3.4cm}|m{1.6cm}|m{1.6cm}| }
 \hline
 Bell state & Eigenvalue  $\sigma_x^{(1)} \otimes \sigma_x^{(2)}$ & Eigenvalue  $\sigma_z^{(1)} \otimes \sigma_z^{(2)}$ \\ 
 \hline
 $\ket{\phi^{+}}=\frac{1}{\sqrt{2}}\left(\ket{00}+\ket{11}\right)$ & +1 & +1 \\ 
\hline
 $\ket{\phi^{-}}=\frac{1}{\sqrt{2}}\left(\ket{00}-\ket{11}\right)$ & $-1$ & +1 \\ 
 \hline
 $\ket{\psi^{+}}=\frac{1}{\sqrt{2}}\left(\ket{01}+\ket{10}\right)$ & +1 & $-1$ \\ 
 \hline
 $\ket{\psi^{-}}=\frac{1}{\sqrt{2}}\left(\ket{01}-\ket{10}\right)$ & $-1$ & $-1$ \\ 
 \hline

\end{tabular}
 \caption{Eigenvalues of the bell state with operators   $\sigma_x^{(1)} \otimes \sigma_x^{(2)}$ and  $\sigma_z^{(1)} \otimes \sigma_z^{(2)}$ }
\label{table:1}
\end{table}

Applying any one channel by itself will obviously pump states to a mixture of the states in the corresponding $-1$ eigenspace (which is \{$\ket{\phi^{-}}$,$\ket{\psi^{-}}$\} for the XX pump and \{$\ket{\psi^{+}}$,$\ket{\psi^{-}}$\} for the ZZ pump).  The forms of the XX pump and ZZ pump are given (following the notation and results of \cite{garcia2020ibm}) as $  \Phi_{xx} \rho_s = E_{1x} \rho_s E_{1x}^{\dagger} + E_{2x} \rho_s E_{2x}^{\dagger} $ and $ \Phi_{zz} \rho_s = E_{1z} \rho_s E_{1z}^{\dagger} + E_{2z} \rho_s E_{2z}^{\dagger}$ with,

\begin{equation}
    E_{1z} = \sqrt{p} \mathbb{I}^{(1)} \otimes \sigma_x^{(2)} \frac{1}{2} (\mathbb{I}+\sigma_z^{(1)} \otimes \sigma_z^{(2)})
    \label{eq:operator_pump_1}
\end{equation}
\begin{equation}
    E_{2z} = \frac{1}{2} (\mathbb{I}-\sigma_z^{(1)} \otimes \sigma_z^{(2)}) +  \frac{\sqrt{1-p}}{2} (\mathbb{I}+\sigma_z^{(1)} \otimes \sigma_z^{(2)})
    \label{eq:operator_pump_2}
\end{equation} 
with $0 \leq p \leq 1$, and $E_{1x}$ and $E_{2x}$ are the same with the replacements $\sigma_z^{(2)} \rightarrow \sigma_x^{(2)}$ and $\sigma_z^{(1)} \otimes \sigma_z^{(2)} \rightarrow \sigma_x^{(1)} \otimes \sigma_x^{(2)}$. 

Given a Bell state, we can map it to a factorised state where one qubit is an eigenstate of $\sigma_z$ and the other is an eigenstate of $\sigma_x$ by applying a CNOT between them. Applying the CNOT on the first system qubit with the second one as control gives us the table \ref{table:2}.

\begin{table}[h!]
\centering
\begin{tabular}{ |c|c| } 
 \hline
 Bell state & Mapped state \\ 
 \hline
 $\ket{\phi^{+}}=\frac{1}{\sqrt{2}}\left(\ket{00}+\ket{11}\right)$ & $\ket{0}\ket{+}$ \\ 
\hline
 $\ket{\phi^{-}}=\frac{1}{\sqrt{2}}\left(\ket{00}-\ket{11}\right)$ & $\ket{0}\ket{-}$ \\ 
 \hline
 $\ket{\psi^{+}}=\frac{1}{\sqrt{2}}\left(\ket{01}+\ket{10}\right)$ & $\ket{1}\ket{+}$ \\ 
 \hline
 $\ket{\psi^{-}}=\frac{1}{\sqrt{2}}\left(\ket{01}-\ket{10}\right)$ & $\ket{1}\ket{-}$ \\ 
 \hline
\end{tabular}
  \caption{Mapping of the bell state }
\label{table:2}
\end{table}

Notice how we can now clearly tell which states need to be pumped. We'll use this to our advantage. 


\subsubsection{ZZ pump}

\begin{figure}[h]
    \centering
    \includegraphics[width=1\textwidth]{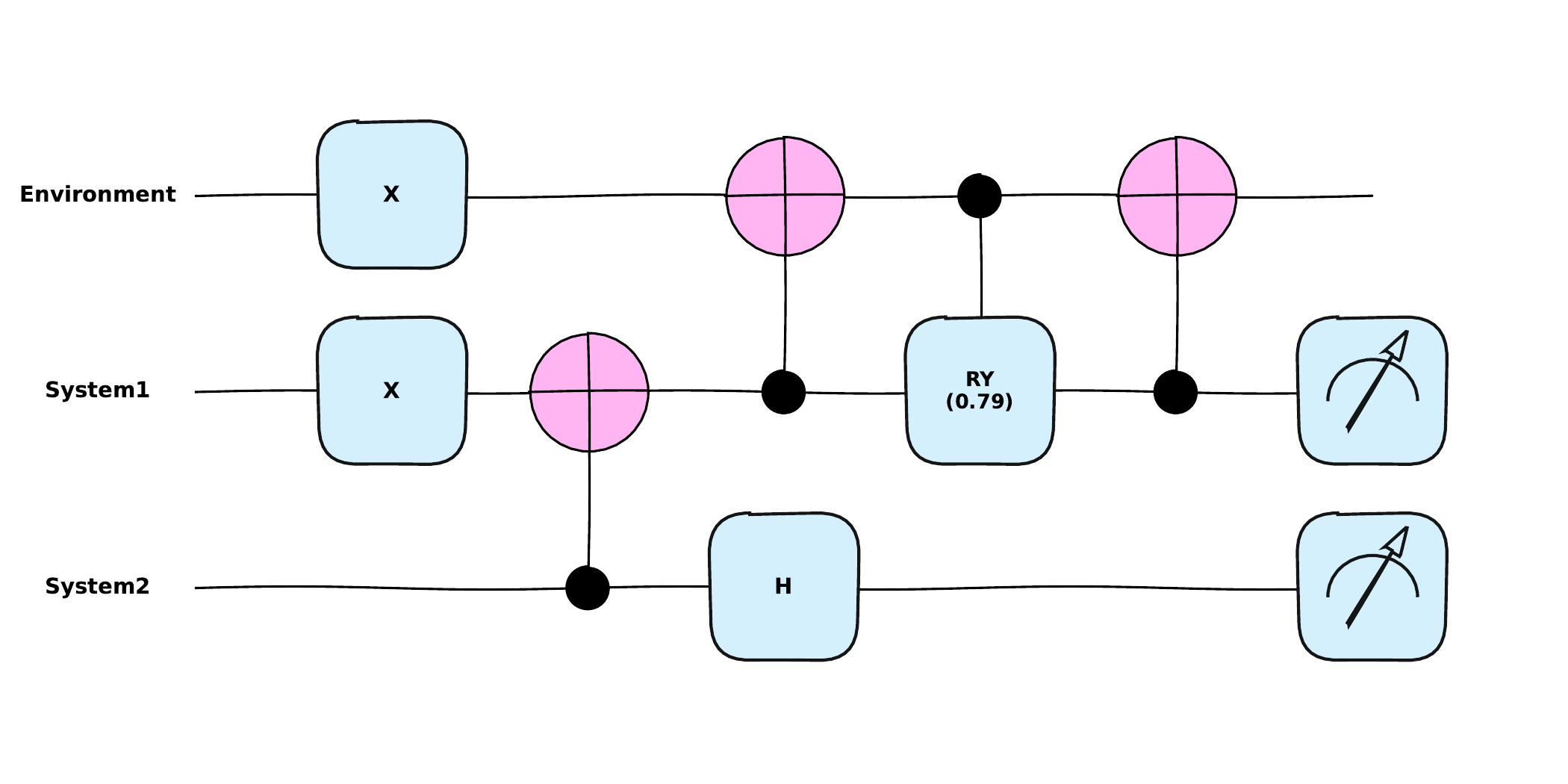}
    \caption{Quantum circuit for the ZZ pump with inital state $\ket{10}$. With $[q1,q2]$ as the system qubits and $q0$ as the environment ancillae. With the angle of the control RY rotation depending on the parameter p.}
    \label{fig:zz_pump_circuit}
\end{figure}

On operating the ZZ pump on the two qubits, we want the populations of $\ket{\psi^+}$ and $\ket{\psi^-}$ to increase since they belong to the $-1$ eigenspace of the $\sigma_z^{(1)} \otimes \sigma_z^{(2)}$ operator.

The pump works by applying an X rotation on the qubit which is an eigenstate of the $\sigma_z$ operator. If the initial Bell state belonged to the $+1$ eigenspace, we will rotate the state to the $-1$ eigenspace with some probability $p$ (the parameter mentioned in Eq. \ref{eq:operator_pump_1} and \ref{eq:operator_pump_2}) using a controlled Y rotation with the angle $\theta=\arccos{(1-2p)}$. If the initial Bell state belonged to the $-1$ eigenspace, we will leave it as it is.

The simplest way to do that is to initialise an enviroment qubit to $\ket{1}$, then apply a CNOT on it controlled by the relevant system qubit, and then apply the pumping X rotation controlled by the environment qubit.
The final circuit for is shown in figure \ref{fig:zz_pump_circuit}

The XX pump follows the same principles as the ZZ pump, but we are dealing with $\ket{+}$ and $\ket{-}$ here instead of $\ket{0}$ and $\ket{1}$. So we first apply a Hadamard gate and then we can do the same thing as before. 


\subsubsection{ZZ-XX pump}
\begin{figure}[h]
    \centering
    \includegraphics[width=1\textwidth]{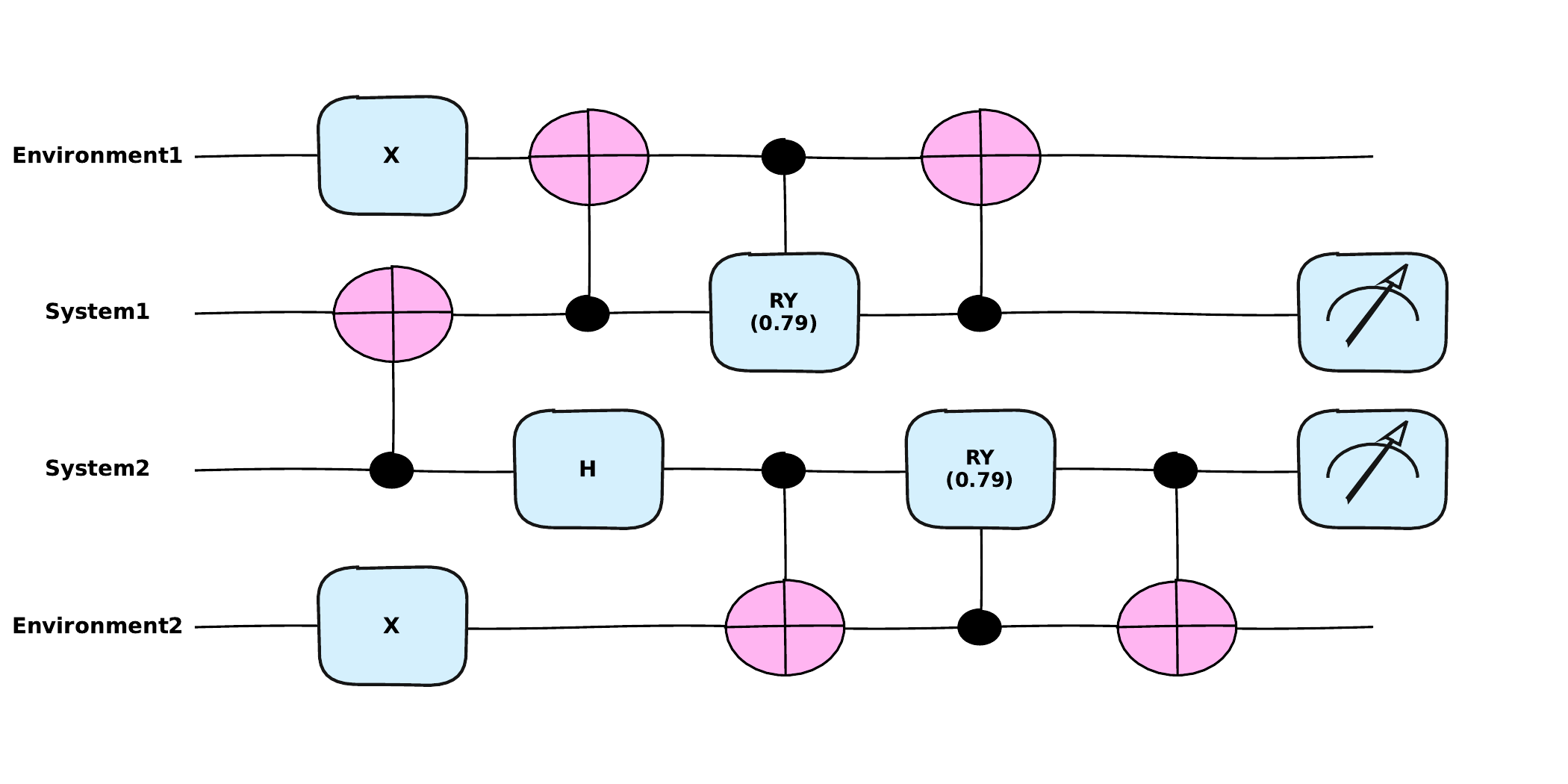}
    \caption{Quantum circuit for the ZZ-XX pump with inital state $\ket{00}$. With $[q1,q2]$ as the system qubits and $[q0,q3]$ as the environment ancillae.With the angle of the control RY rotation depending on the parameter p.}
    \label{fig:zz_xx_pump_circuit}
\end{figure}

We will concatenate the ZZ and XX pumps and remove the final CNOT of the ZZ pump and the first CNOT of the XX pump. The final circuit for is shown in figure \ref{fig:zz_xx_pump_circuit}

\begin{figure}[h]
    \centering
    \includegraphics[width=0.95\textwidth]{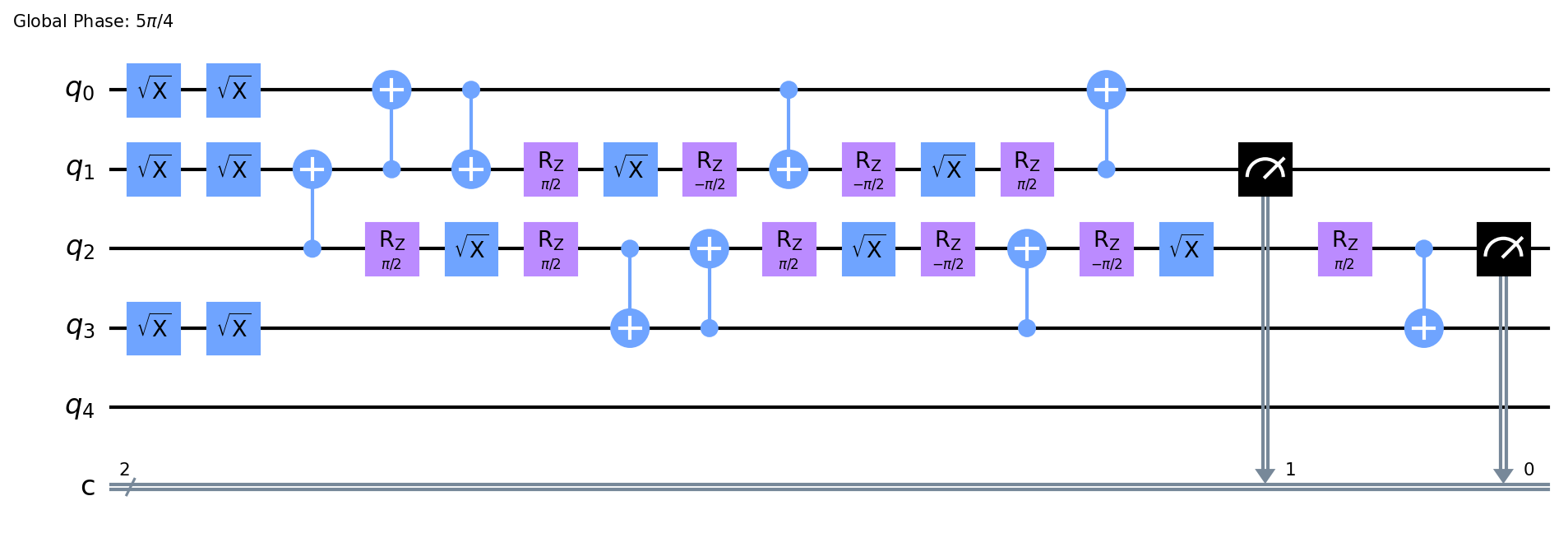}
    \caption{Transpiled quantum circuit for the ZZ-XX pump with inital state $\ket{10}$. With gate basis $[cx, id, rz, sx]$}
    \label{fig:zz_xx_pump_transpiled_circuit}
\end{figure}

In figure \ref{fig:zz_xx_pump_transpiled_circuit} we can observe the transpiled circuit for the ZZ-XX pump with the gate basis used in the IBM devices, $[cx, id, rz, sx]$. We can observe the increase in depth and in the number of CNOTs used. 


\subsection{Collisional model}

This model was developed as the Markovian reservoir by \cite{garcia2020ibm}, following the work presented in \cite{filippov2017divisibility}. In this model, a collision consists of applying the unitary operator $U = e^{ig\tau\sigma_z} \otimes \ket{0}_k\bra{0} + e^{-ig\tau\sigma_z} \otimes \ket{1}_k\bra{1}$ between the system qubit and the $k^{th}$ environmental qubit. Here, $g$ is the coupling strength and $\tau$ is the time duration over which one collision takes place and $n$ collisions occurs in the time $t=n\tau$. Now we consider two cases. 


\subsubsection{Correlated case}

We'll take our environment qubits prepared in the classically correlated state $\rho = \frac{1}{2}(\ket{0}^{\otimes n}\bra{0}^{\otimes n}+\ket{1}^{\otimes n}\bra{1}^{\otimes n})$. If we consider $\rho_S = a\ketbra{0}{0}+b\ketbra{0}{1}+c\ketbra{1}{0}+d\ketbra{1}{1}$, then, $a=b=c=d=\frac{1}{2}$.

While density matrix elements can be complex in general, in this case it is enough to only get the real part of $\rho_{12}$.

The dynamical map after $n$ collisions (following the results of \cite{garcia2020ibm}) is given by

\begin{equation}
\begin{aligned}
    \Phi_t \rho_S & = Tr_E [U_n \cdots U_2 U_1 (\rho_S \otimes \rho_{corr}) U_1^\dag U_2^\dag \cdots U_n^\dag] \\ 
    &= \cos^2{(ng\tau)} \rho_S + \sin^2{(ng\tau)} \sigma_z \rho_S \sigma_z
\end{aligned}
\end{equation}

(since there is only one system qubit), then applying the dynamical map above gives us

\begin{equation}
\begin{aligned}
    \Phi_t \rho_S & = a\ketbra{0}{0}+(\cos^2{(ng\tau)}-\sin^2{(ng\tau)})b\ketbra{0}{1}+ \\  & (\cos^2{(ng\tau)}-\sin^2{(ng\tau)})c\ketbra{1}{0}+d\ketbra{1}{1}
\end{aligned}
\end{equation}

Only affecting the coherences, and it multiplies the same factor to both. Given this, only need to check $\rho_{12}$ in order to quantify the effect of the dynamical map.

In the general structure of each circuit, first, the system qubit is initialised to $\ket{+}$ using a Hadamard gate. Then we prepare the environment qubits, a series of $i$ collisions follow where $i \in [1,c]$. Finally, we apply a Hadamard gate on the system qubit, with the following effect,

\begin{equation}
\begin{aligned}
    H \rho_S H^{\dagger}
    & =
    \frac{1}{\sqrt{2}}
    \begin{bmatrix}
        1 & 1 \\
        1 & -1
    \end{bmatrix}
    \begin{bmatrix}
        \frac{1}{2} & \frac{x}{2} \\
        \frac{x}{2} & -\frac{1}{2}
    \end{bmatrix}
    \frac{1}{\sqrt{2}}
    \begin{bmatrix}
        1 & 1 \\
        1 & -1
    \end{bmatrix} \\
    &=
    \begin{bmatrix}
        \frac{1+x}{2} & 0 \\
        0 & \frac{1-x}{2}
    \end{bmatrix}
\end{aligned}
\end{equation}
We can then measure the populations of $\ket{0}$ and $\ket{1}$, subtract them, and then divide by number of shots to find out $x$ (where $x$ is the factor the coherences are multiplied by in each case). Then we know that $\rho_{12}=\frac{x}{2}$.

A collision is performed in the following way. We apply a CNOT gate on the system qubit controlled by the relevant environment qubit, then apply a Z rotation by the required factor, then apply the same CNOT again. 

\begin{figure}[h]
    \centering
    \includegraphics[width=1\textwidth]{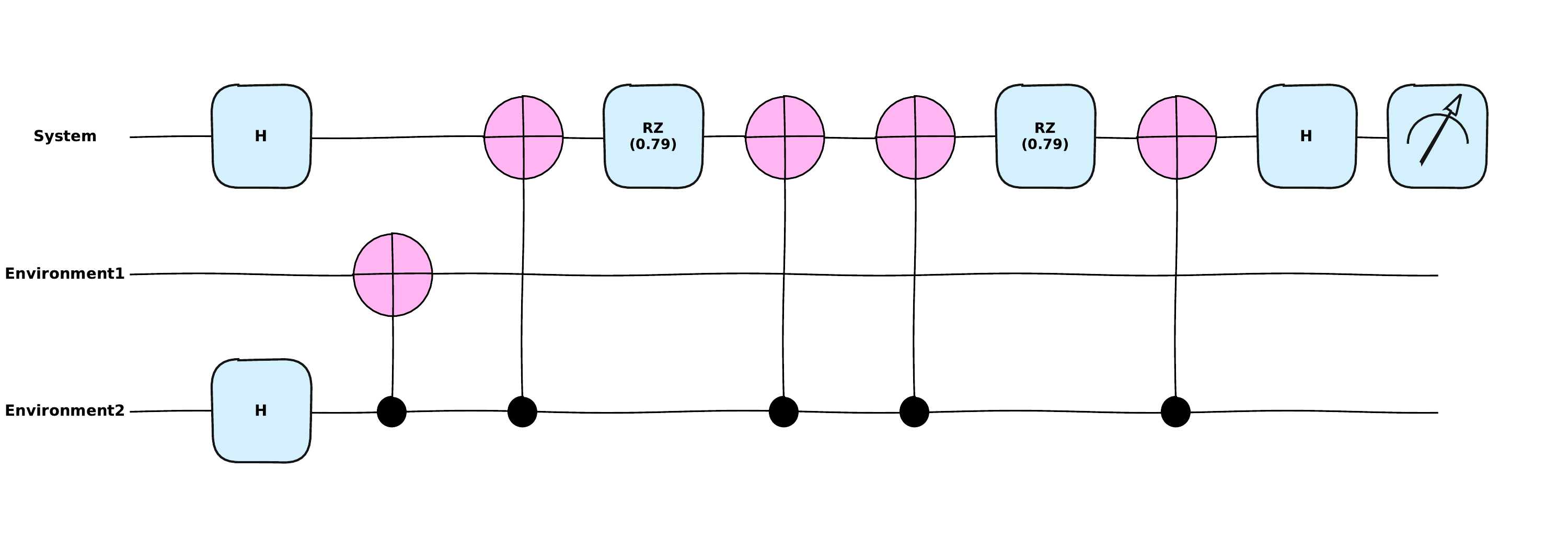}
    \caption{Circuit diagram for 2 collisions in the correlated case.}
     \label{fig:correltaed_circuit}
\end{figure}

In order to prevent the compiler from optimizing the circuit in an undesired way, we have instructed it to run the circuit without any further gate optimization through the transpiler. We used the verbatim feature in AWS Braket and the Qiskit tranpiler options. This is necessary because the compiler may remove one of the CNOT gates that are involved in each collision, and combine them into a single rotation gate, which is not what we want to achieve.


\subsubsection{Uncorrelated case}

In this case, the environment qubits are prepared in the uncorrelated state $\ket{+}^{\otimes n}$. The dynamical map after $n$ collisions (following the results of \cite{garcia2020ibm})  is given by

\begin{equation}
    \Phi_t \rho_S = \frac{1}{2}(1+\cos^n{(2g\tau)}) \rho_S + \frac{1}{2}(1-\cos^n{(2g\tau)}) \sigma_z \rho_S \sigma_z
\end{equation}

Like before, considering $\rho_S = a\ketbra{0}{0}+b\ketbra{0}{1}+c\ketbra{1}{0}+d\ketbra{1}{1}$ gives us

\begin{equation}
\begin{aligned}
    \Phi_t \rho_S  & = a\ketbra{0}{0}+\cos^n{(2g\tau)}b\ketbra{0}{1}+ \\ 
     & \cos^n{(2g\tau)}c\ketbra{1}{0}+d\ketbra{1}{1}
\end{aligned}
\end{equation}

The form of the result is similar as the correlated case, so the same conclusions follow.

\begin{figure}[h]
    \centering
    \includegraphics[width=1\textwidth]{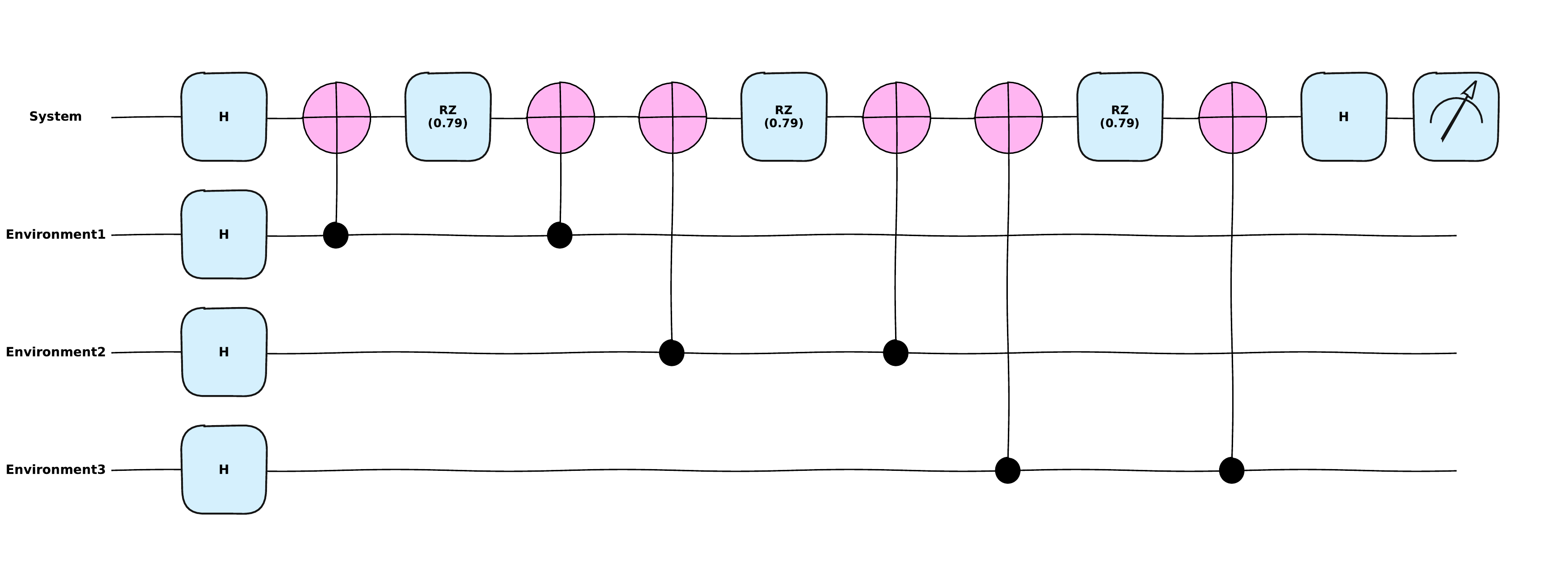}
    \caption{Circuit diagram for 3 collisions in the uncorrelated case.}
    \label{fig:uncorreltaed_circuit}
\end{figure}

In this case, we simply apply a Hadamard gate on each environment qubit to get them to the state $\ket{+}^{\otimes n}$. Then we perform collisions with each of the environment qubits sequentially. 
The circuit is shown in figure \ref{fig:uncorreltaed_circuit}.


\section{Noise models}

A method to study the behavior of Quantum Circuits before running on real quantum computers is to create a noise model. It can be consider that some of the main errors in a QPU are decoherence and gate errors \cite{georgopoulos2021modeling}. In this work, we focused on a noise model with different fidelities of the CNOT gate and one qubit gates. Decreasing the fidelity will affect the results considerably. Finding a reasonable fidelity in the simulations will give us an estimate of the resources we will need for a perfect outcome of the circuits in real QPU. For QPUs from
IBM, taking the information from the IBM quantum platform for the \texttt{ibmq\_peekskill} at 13 of october of 2023. The single-qubit gates’ error rates are approximately $10^{-4}$ ( $\approx 99.98 \%$) and the two-qubit CNOT gate’s error rate is $10^{-3}$ ( $\approx 99.41 \%$). This errors could change depending on the day and the device.

In the noise models used a simple model with errors from 1 and 2 qubit gate, which was modeled by applying depolarising noise. For the real hardware we used the latest devices from IBM with 127 qubits, \texttt{ibmq\_osaka} and \texttt{ibmq\_tokyo}, and the OQC device available trough the \texttt{qBraid} environment in AWS. 


\section{Quantum Error Mitigation}

We have applied Quantum Error Mitigation using the Zero-Noise Extrapolation (ZNE)\cite{giurgica2020digital, temme2017error} technique from the Mitiq toolkit \cite{larose2022mitiq}. ZNE extrapolate the expectation value with zero noise of an observable from a list of expectation values computed at different noise scales. In this case, the extrapolation is done over the measure state overlap, different extrapolations methods were tested with a minimum of 4 scale factors and a maximum of 8. To scaled the noise the unitary fold gates at random method has been used, which performs the transformation $G \rightarrow GG^\dagger G$ where $G$ is some gate (or to a random subset of gates). 
The unitary folding approach has a major advantage that distinguishes it from other methods. It is entirely digital, which means that the noise can be scaled using a high level of abstraction from the physical hardware. Furthermore, this approach does not require knowledge of the intricate details of the underlying noise model. This makes it an attractive option for a reliable solution that is easy to implement.

Readout Error Mitigation (REM) \cite{bravyi2021mitigating} was implemented as well. Errors that occur during the measurement of qubits in a quantum device can negatively impact quantum computation. These errors are known as readout errors. By thoroughly analyzing and understanding the readout errors, it is possible to create a readout error mitigator. This tool can be used to obtain more accurate output distributions, as well as more precise measurements of expectation values for measurables. In this work we used the REM technique with the confusion matrix from Mitiq, a square matrix that encodes the measurement error for each pair of basis states. 

Comparison of the results for the mitigated and unmitigated cases are shown for each model. The code to reproduce this results are in the Github repository \footnote{\url{https://github.com/JessicaJohnBritto/QOSF_Cohort6}}


\section{Results}

\subsection{Simulation Markovian reservoir}

\begin{figure}[h]
      \includegraphics[width=0.9\linewidth]{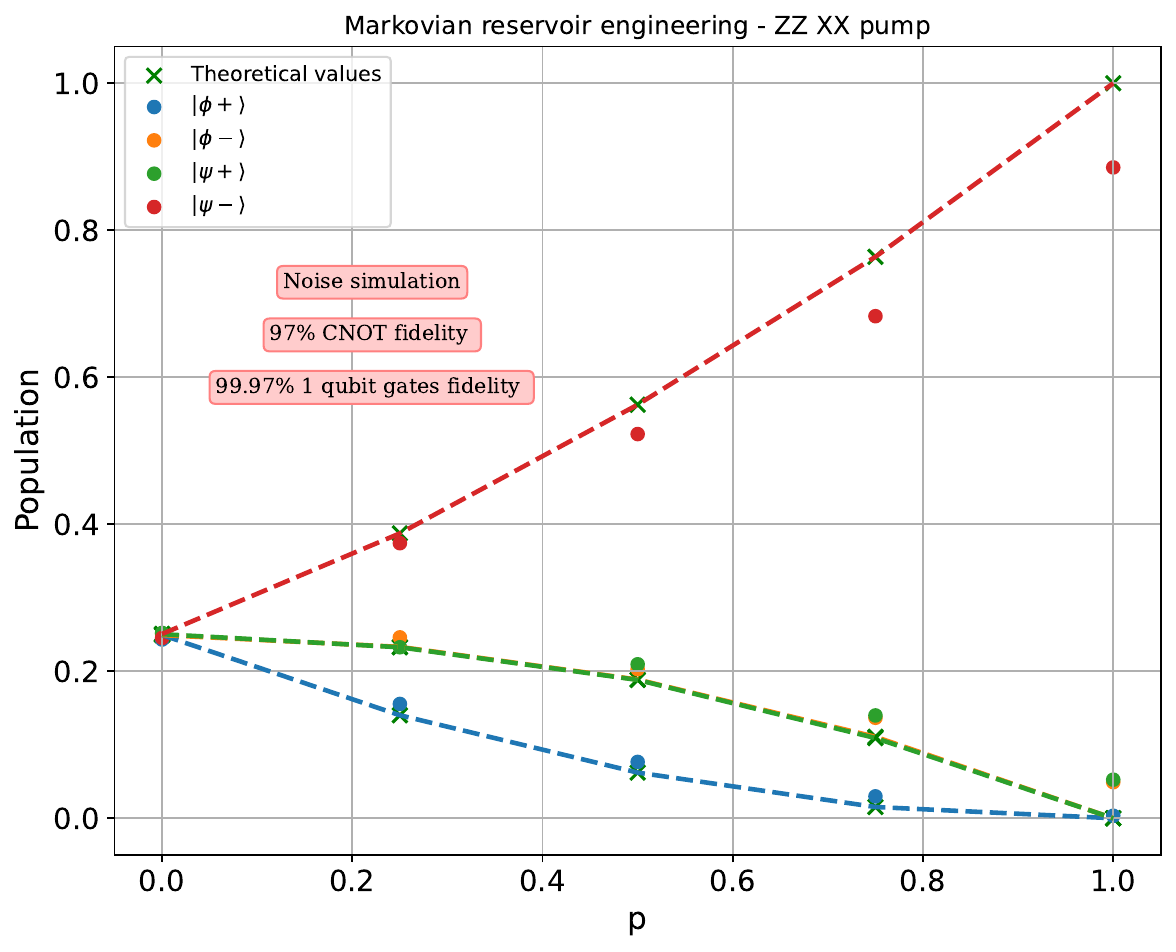} 
      \includegraphics[width=0.9\linewidth]{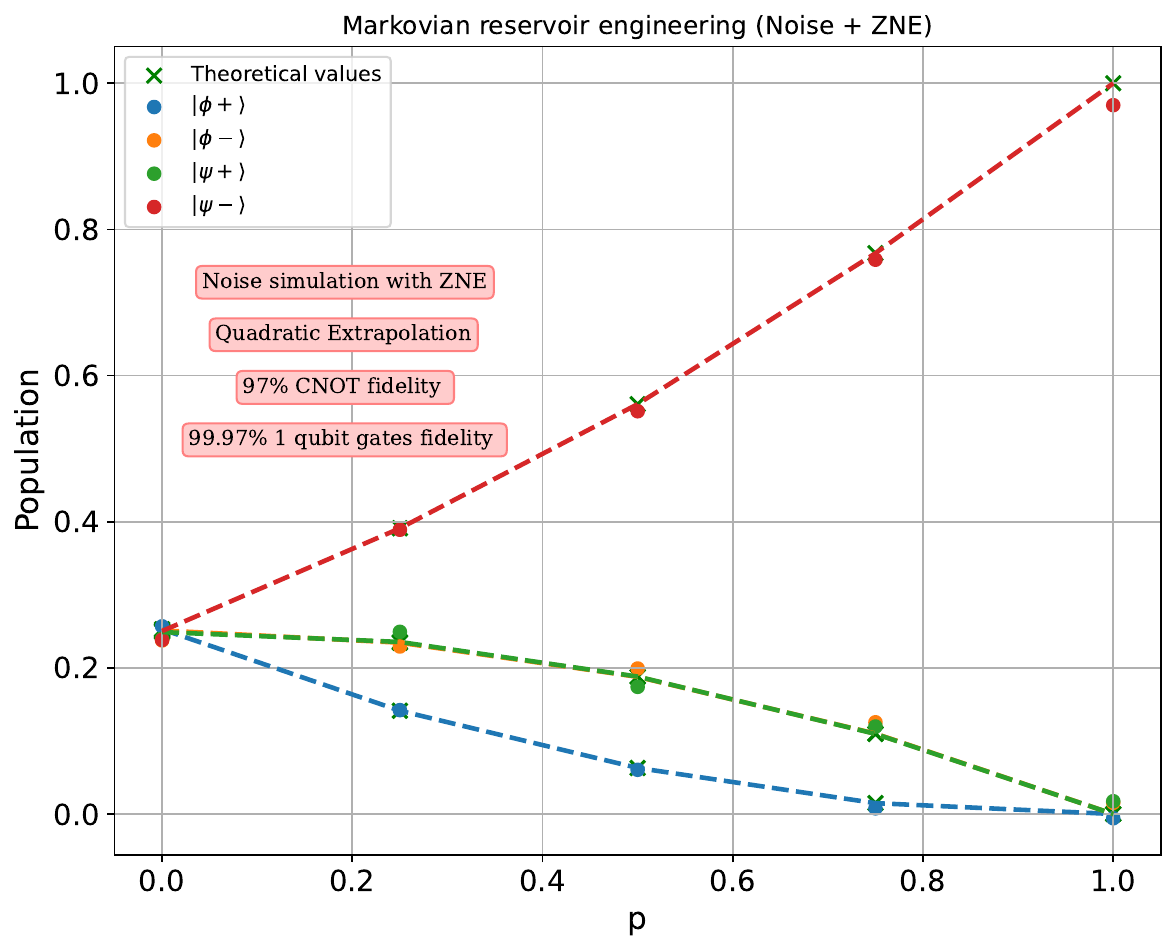} 
  \caption{(Top) Results with noise simulation for the ZZ-XX pump with fidelity of CNOT gates of 97 \% and the one qubit gates of 99.97 \%. (Bottom) Results with ZNE using a linear extrapolation with 4 scaling factors. The dashed line is the analytical prediction.In both cases 1024 shots is used. }
  \label{fig:noisy_markovian}
\end{figure}

\begin{figure}[h]
    \centering
    \includegraphics[width=1\textwidth]{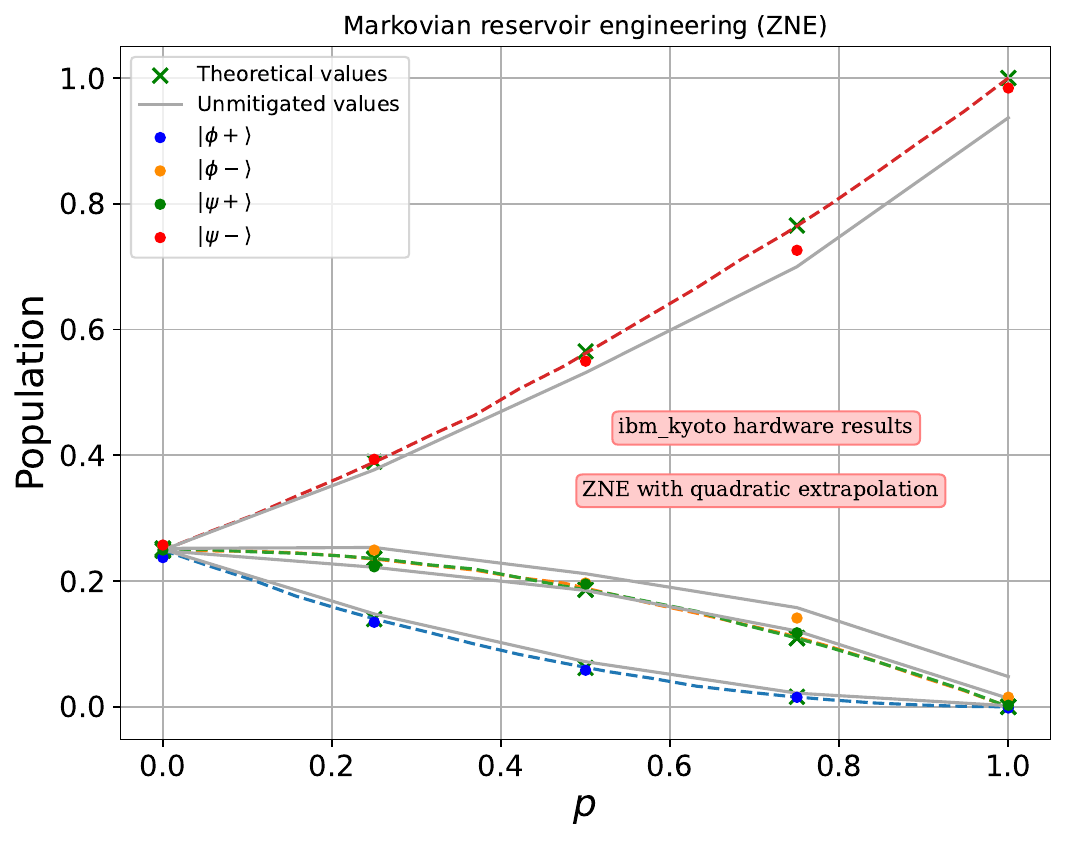}
    \includegraphics[width=1\textwidth]{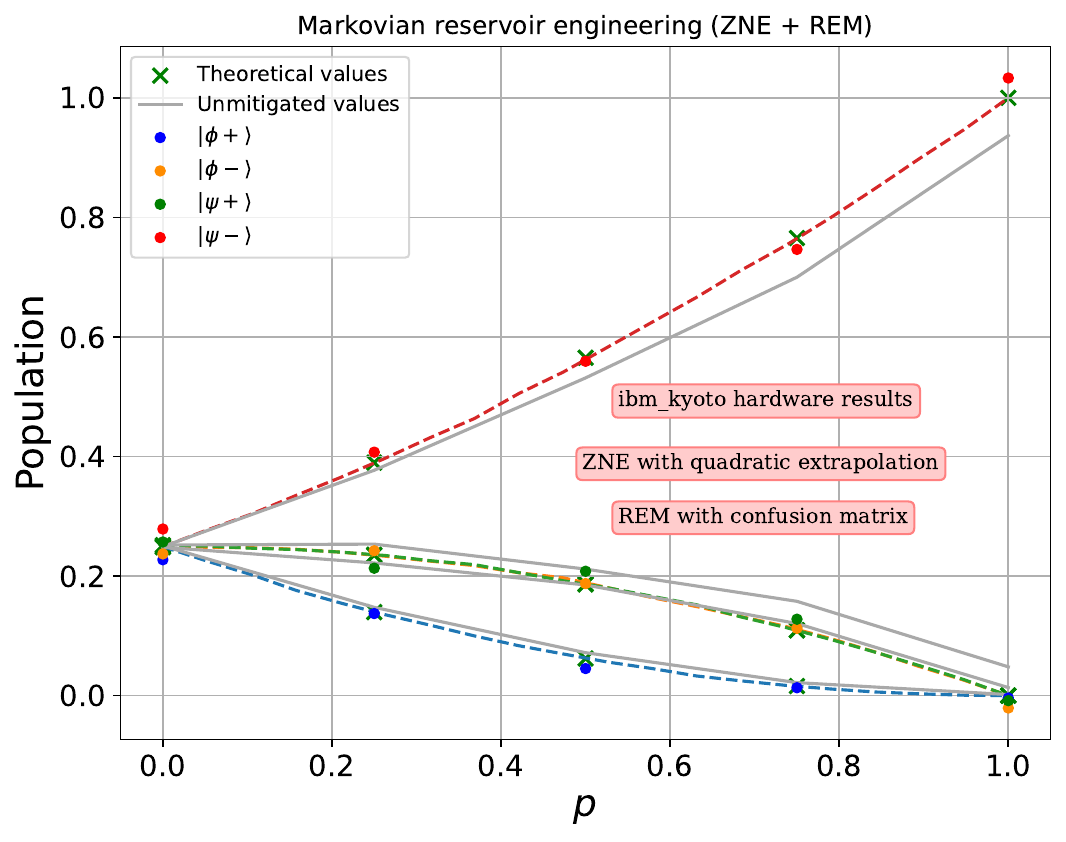}
    \caption{Results from \texttt{ibmq\_kyoto} for the ZZ-XX pump (Top) Results with ZNE with quadratic extrapolation and 1024 shots. (Bottom) Results with ZNE and REM using the confusion matrix method. The points correspond to the experimental results, while the dashed lines show the theoretical prediction.  }
     \label{fig:results_markovian}
\end{figure}

In figure \ref{fig:noisy_markovian} we can see the results of the Markovian reservoir for the ZZ-XX pump with noisy simulations. Also, we can observe the improvements at the moment of using ZNE. Each plot compares the overlap between the state of the system and the Bell states (dots in figure \ref{fig:noisy_markovian} and \ref{fig:results_markovian}).

In figure \ref{fig:results_markovian} we can observe the results of running the simulation in a real quantum device, \texttt{ibmq\_kyoto}, for the  ZZ-XX pump where the population of states $\ket{\psi^-}$ are clearly increasing with $p$. To compare the results of the OQC device Lucy, we run the same circuits, taking into account Lucy's topology, particularly the direction of two-qubit gates. In the case  of the states $\ket{\phi^+}$ and $\ket{\phi^-}$ are decreasing, but slowly, because their populations are increased by one pump and decreased by the other. The population of state $\ket{\phi^+}$ decreases the fastest because its population is decreased by both pumps. ZNE is applied with a quadratic extrapolations and scale factors of [1, 3, 5, 7].

\begin{figure}[h]
    \centering
    \includegraphics[width=1\textwidth]{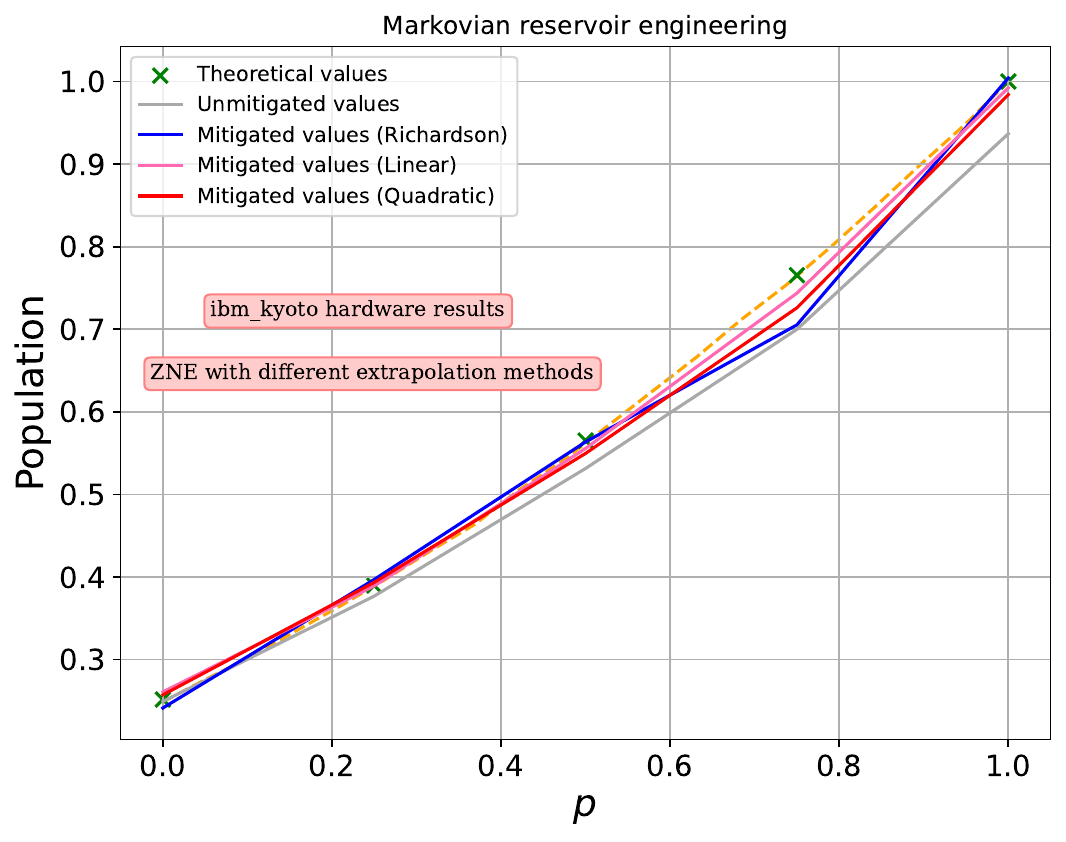}
    \caption{Results from \texttt{ibm\_kyoto} (Left) Mitigated Results using ZNE with Mitiq. Different Extrapolations - Richardson, Linear and Quadratic for $\ket{\psi^-}$.}
     \label{fig:MRE_different_extrapolations_1024}
\end{figure}

In figure \ref{fig:MRE_different_extrapolations_1024}, we used different extrapolations that are available in Mitiq - Richardson, Linear and Quadratic, in an attempt to find out which extrapolation gives mitigated results that are closest to the theoretical results. This was particularly done for the case of $\ket{\psi^-}$. And, for the same case, we used scale factors of [1, 3, 5, 7]. The points in figure \ref{fig:MRE_different_extrapolations_1024} correspond to the experimental results with 1024 shots, while the dashed lines show the theoretical prediction. From figure \ref{fig:MRE_different_extrapolations_1024}, it is clear that Quadratic Extrapolation gives the results closest to the theoretical results. In the case of the noisy simulations, the best extrapolation method was the linear. This difference can be attribute at the complexity of the noise in a real quantum device compare with the depolarising noise simulations used.  

It is important to highlight the improvements on previous works, in \cite{garcia2020ibm} for the Markovian reservoir was necessary to use 8192 shots to obtain accurate results. In our case we observe that with 1024 shots is enough for satisfactory results, even though, at the moment of using ZNE the circuits are being run multiple times. In \cite{abu2023preparing}, the number of shots and error mitigation technique was not specified. 

\FloatBarrier

\begin{figure}[h]
    \centering
    \includegraphics[width=1\textwidth]{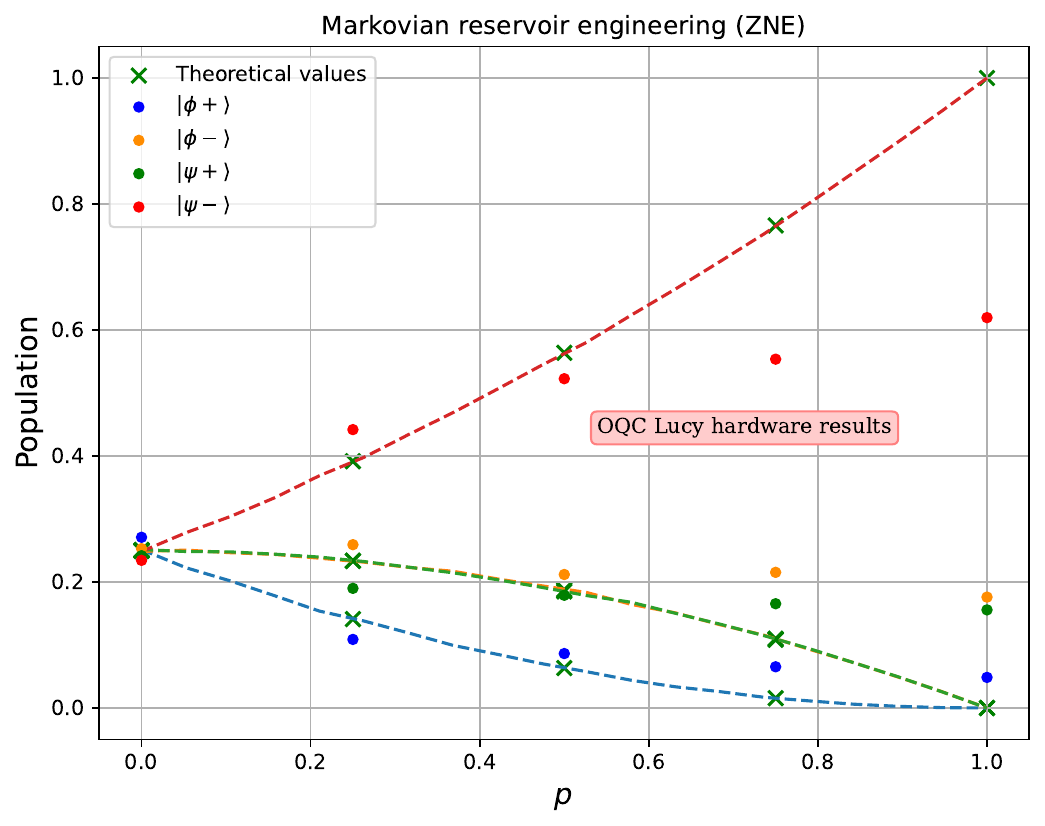}
    \caption{Results from OQC Lucy for the ZZ-XX pump. The points correspond to the experimental results, while the dashed lines show the theoretical prediction. Given the noise of the system the results are inconsistent. }
     \label{fig:results_markovian_lucy}
\end{figure}

In figure \ref{fig:results_markovian_lucy} we can observe the results of the Markovian model in the Lucy device from OQC. Given the noise of the system the results shown an inconsistent simulation of the model, even though we chose a qubit mapping with the best quality of connections. Applying ZNE to this high noisy results will not improve the outcome.


\subsection{Simulation of the Collisional Model}

\subsubsection{Correlated Case}

\begin{figure}[h]
      \includegraphics[width=0.48\linewidth]{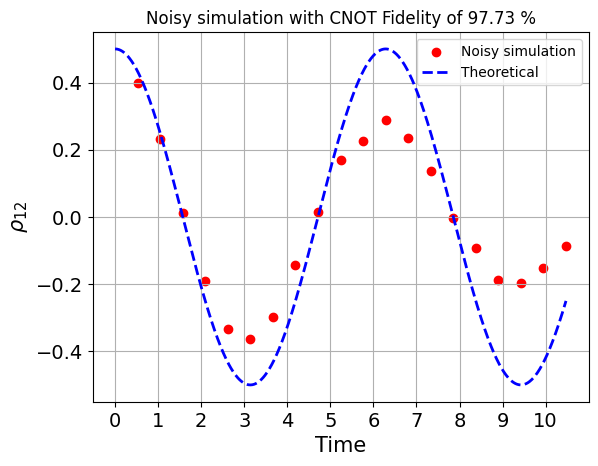} 
      \includegraphics[width=0.48\linewidth]{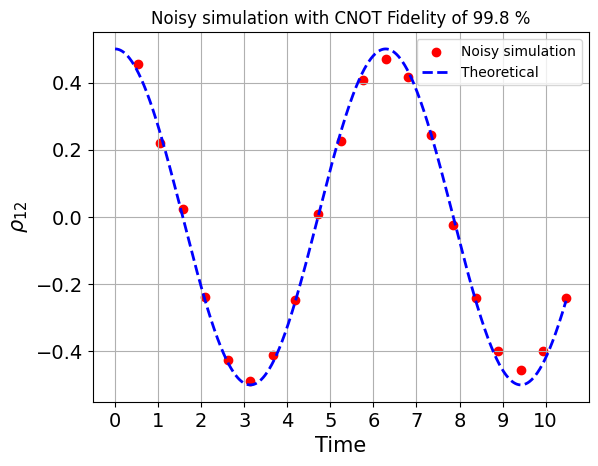} 
  \caption{Results with different fidelity associated with the two-qubit gates with a constant fidelity on the one qubit gates of 99.97 \%. With 1024 shots and 20 correlated collisions}
  \label{fig:correlated_case_fidelity}
\end{figure}

In figure \ref{fig:correlated_case_fidelity} we can see that with a decrease in fidelity, the results deviate from the theoretical curve, such that the system undergoes damping. The region surrounding the maximas and minimas have maximum deviation from the theoretical results.

\begin{figure}[h]
    \centering
    \includegraphics[width=0.9\textwidth]{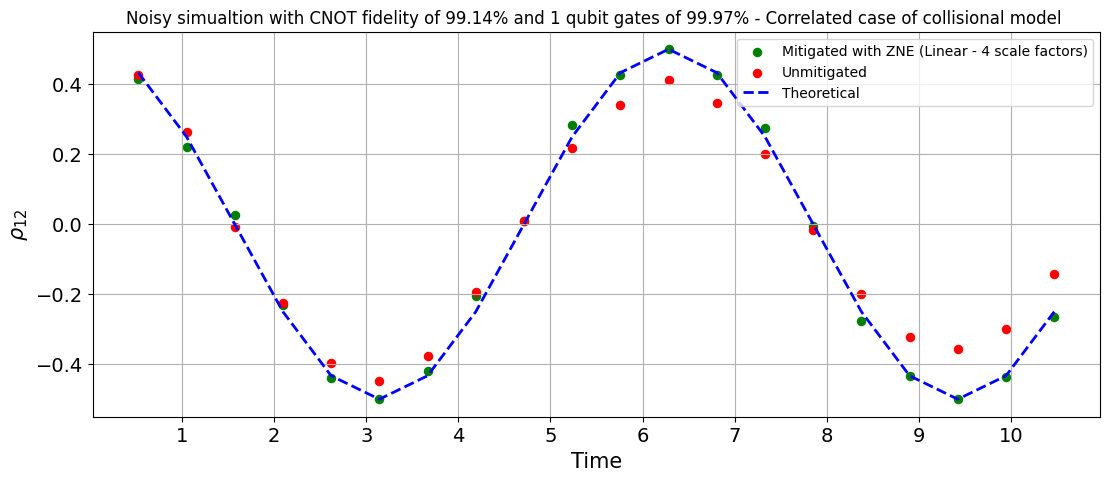}
    \caption{Simulated noisy results (coherences vs time $t$) with 99.14 \% fidelity in the CNOT gates for the collisional model with the correlated case. For the theoretical, mitigated and unmitigated cases using ZNE with 1024 shots and 20 collisions.}
    \label{fig:correlated_case_zne}
\end{figure}
\FloatBarrier

In figure \ref{fig:correlated_case_zne} we can observe the results with noisy simulations using the ZNE technique for error mitigation with a linear extrapolation and 4 scale factors. We can observe a corrections almost perfect with the mitigated results.

\begin{figure}[h]
      \includegraphics[width=0.45\linewidth]{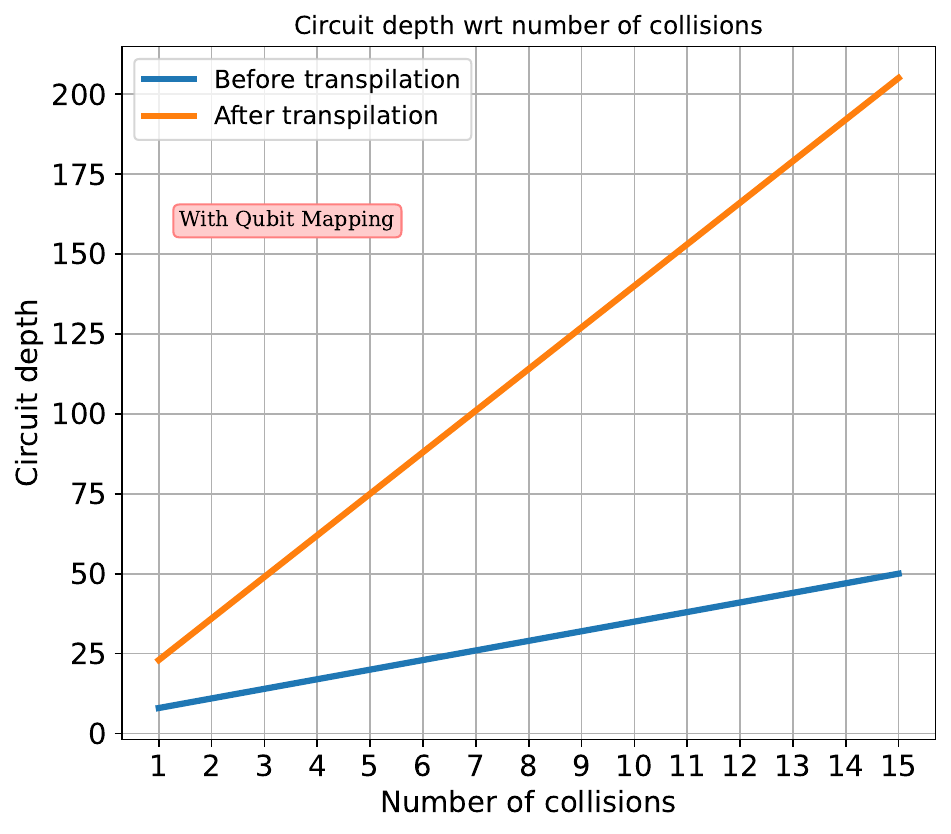} 
      \includegraphics[width=0.45\linewidth]{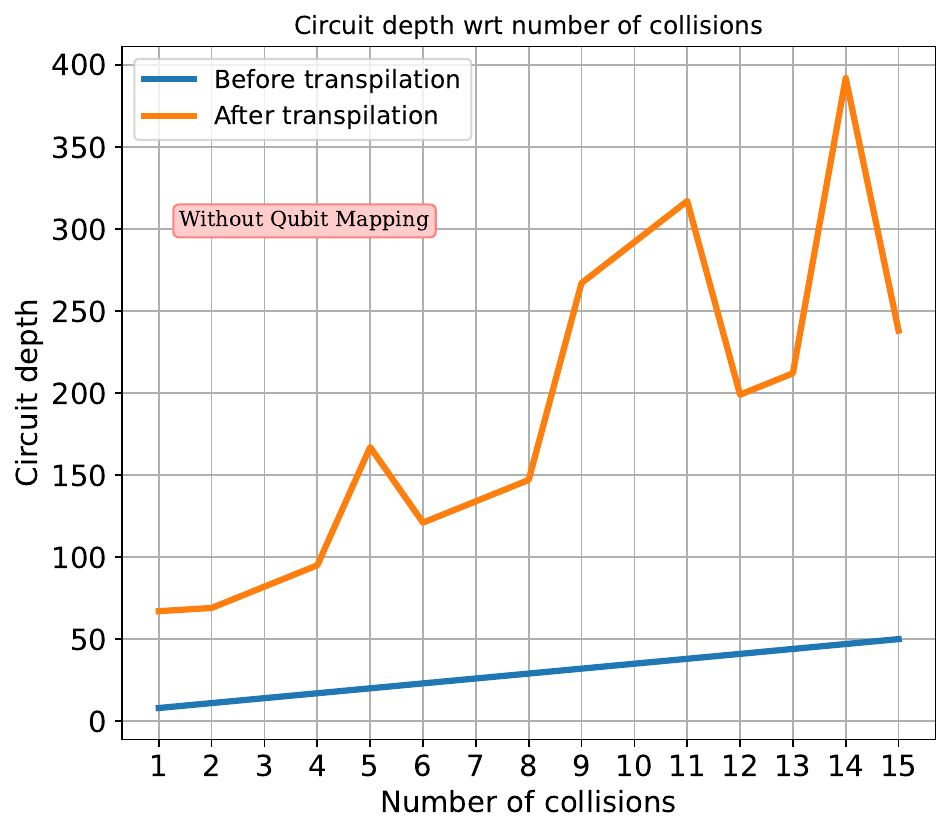} 
  \caption{Circuit Depth w.r.t number of collisions (Left) After Qubit Mapping. (Right) Without using Qubit Mapping}
  \label{fig:qubit-mapping_and_depth}
\end{figure}

Running the collisional model in real quantum hardware is more complicated than the Markovian Reservoir model. The increase in depth of the circuits is linear with the number of collisions. To understand how this affects the circuits we can see in figure \ref{fig:qubit-mapping_and_depth}, the depth with different number of collisions for the case of \texttt{ibm\_kyoto}. If the qubit mapping is not specified the depth of the circuit becomes unstable, in our case we chose a mapping given the best qubits in the calibrations data of IBM Quantum.  For the case of \texttt{OQC\_Lucy}, we used qubits 4 and 5 from the hardware based on their connectivity (as shown in fig.\ref{fig:Architure_QH_Used_lucy}) and fidelity rates. With IBM we ran the same circuits, taking into account Lucy's topology, as with the Markovian reservoir model.

\begin{figure}[h]
      \includegraphics[width=1\linewidth]{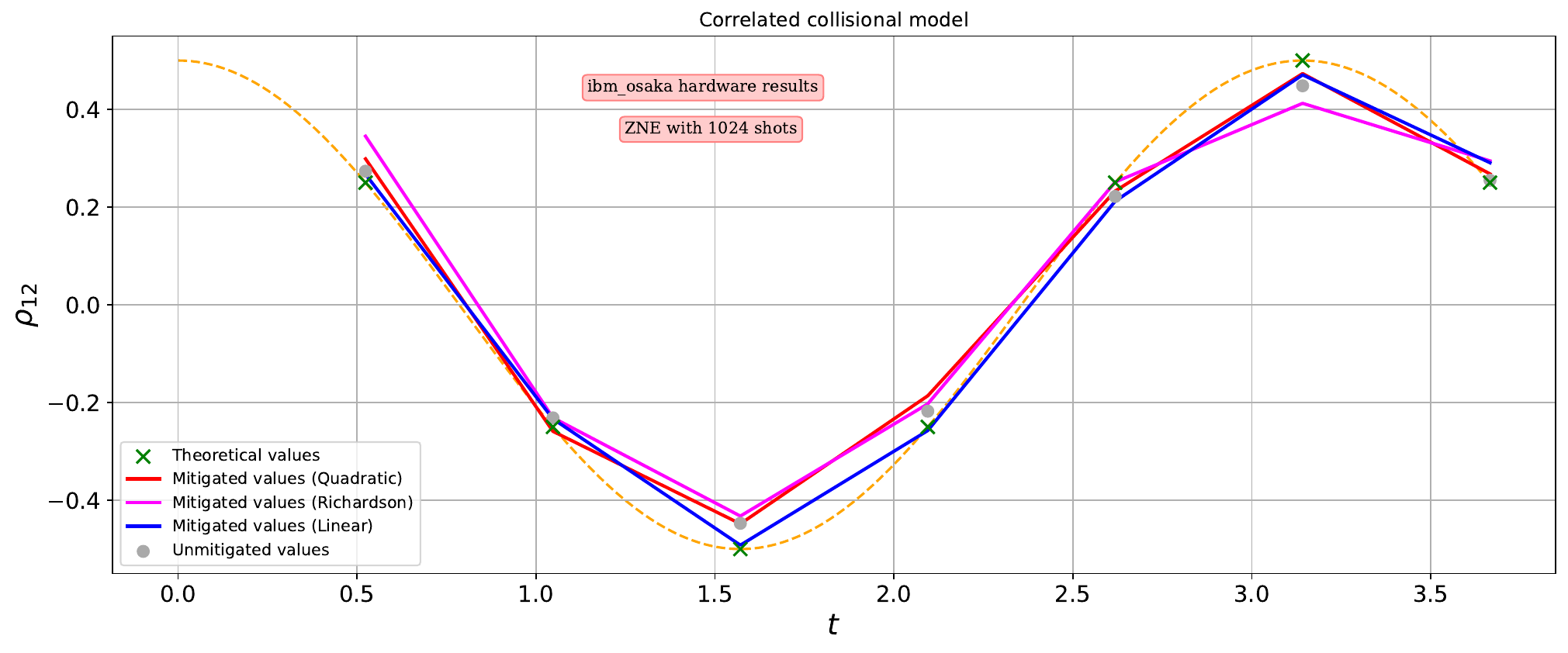} 
  \caption{Results from \texttt{ibmq\_kyoto} for the collisional model in the correlated case. The results are with different extrapolations methods using ZNE. The grey points correspond to the experimental,while the dashed lines show the theoretical prediction.}
  \label{fig:results_ibm_correlated}
\end{figure}

In figure \ref{fig:results_ibm_correlated}, we observe a comparison between different extrapolation methods for scale factors of [1, 3, 5, 7]. The experimental results are from \texttt{ibm\_osaka} and with 1024 number of shots. The linear extrapolations shows the best results. In this case the best extrapolations method with the noise models and the quantum device are the same. Comparing with the results in \cite{garcia2020ibm}, we can observe an improvement in the unmitigated values with only 1024 shots, which is improve even more in certain cases with ZNE. 

\begin{figure}[h]
    \centering
    \includegraphics[width=0.9\textwidth]{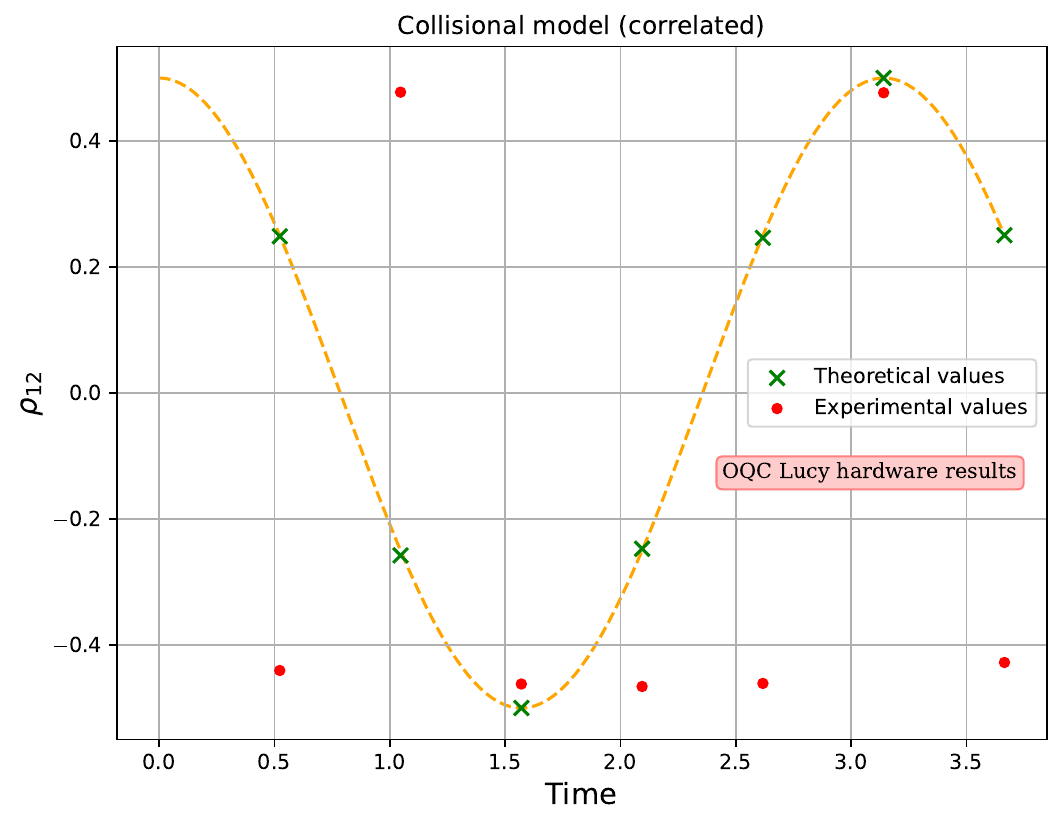}
    \caption{Results from OQC Lucy for the correlated collisional model. In this case we iterate trough two different qubits for each collision. The red points in the graph represent the experimental results, while the dashed lines indicate the theoretical prediction. However, due to the noise present in the system, the results obtained are not consistent. }
     \label{fig:results_collisional_lucy}
\end{figure}

\begin{figure}[h]
    \centering
    \includegraphics[width=0.9\textwidth]{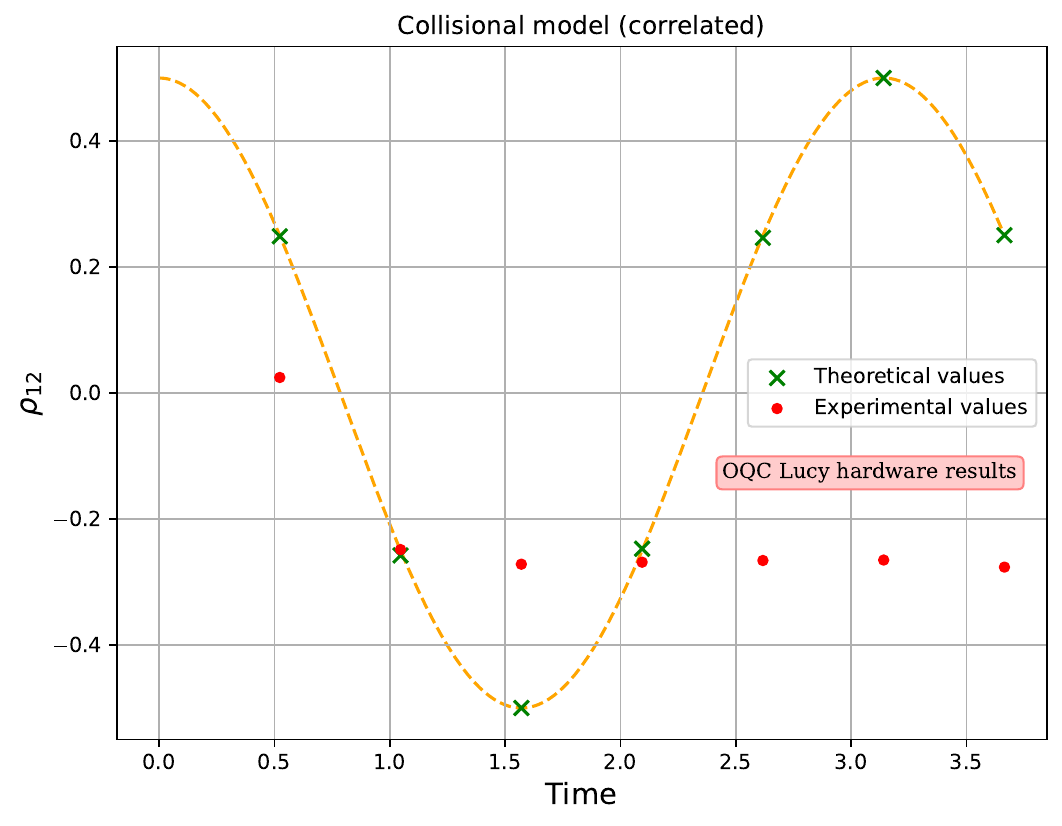}
    \caption{Results from OQC Lucy for correlated collisional model. In this case we using the same circuit shown in figure \ref{fig:correltaed_circuit}, using the same qubit for the different collisions. To avoid unwanted optimization we used the verbatim option in AWS Braket. }
     \label{fig:results_collisional_lucy_verbatim}
\end{figure}

In the figure \ref{fig:results_collisional_lucy_verbatim}, we can see the simulation results of using the circuit shown in Figure \ref{fig:correltaed_circuit}, where we utilized the same qubit for all collisions. To avoid any undesired optimization, we enabled the verbatim feature in AWS Braket, which runs the circuit without any gate optimizations. We also experimented with the option of not using the verbatim feature, and in this case, we changed the circuit to use two different qubits for each collision to prevent any unwanted optimization. The results are shown in the figure \ref{fig:results_collisional_lucy}. However, in both cases, just like with the Markovian model simulations, the system's noise was too high, affecting the simulation output. Each simulation produced inconsistent results, despite selecting an appropriate qubit mapping for the hardware. It is worth noting that when we used four qubits without the verbatim feature, the results began oscillating between similar values. This behavior can be attributed to a uniform type of noise that the quantum device produces.

\subsubsection{Uncorrelated case}

\begin{figure}[h]
      \includegraphics[width=0.45\linewidth]{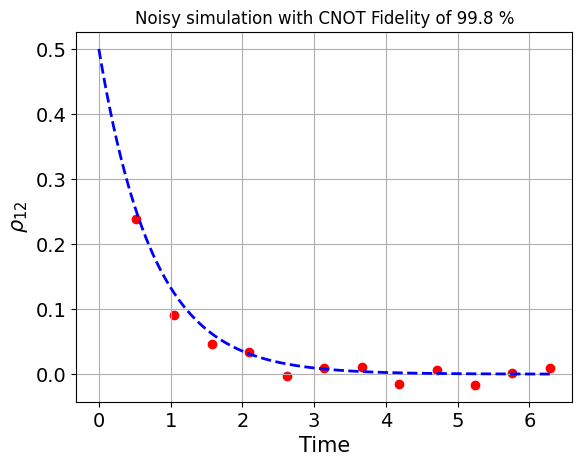}
      \includegraphics[width=0.45\linewidth]{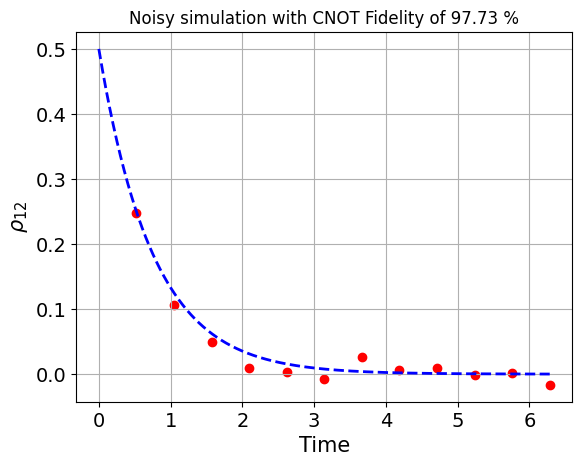} 
  \caption{Results of the noisy simulation of the collisional model for the uncorrelated case with different fidelities of CNOT gates and one fidelity of one qubit gates of 99.97 \% using 1024 shots with 12 collisions.}
\label{fig:uncorrelated_case_noisy_cases}
\end{figure}

In figure \ref{fig:uncorrelated_case_noisy_cases} we can observe the results of the noisy simulations of the uncorrelated collisional model by varying the fidelity associated with the two-qubit gates, the CNOT. In all cases we use 1024 shots with 12 collisions.

\begin{figure}[h!]
    \centering
    \includegraphics[width=1\textwidth]{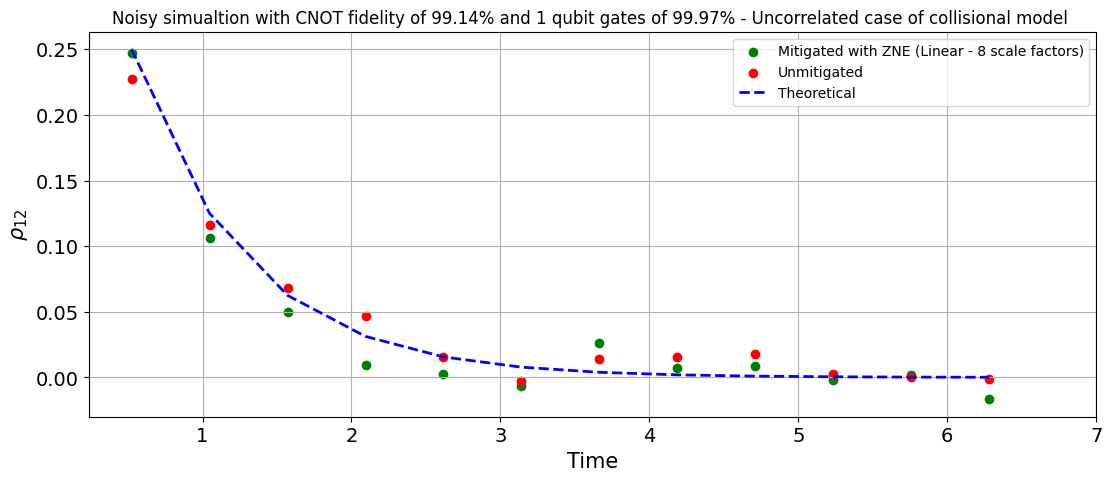}
    \caption{Simulated noisy results (coherences vs time $t$) with 99.14 \% fidelity in the CNOT gates for the collisional model with the uncorrelated case. For the theoretical, mitigated and unmitigated cases using ZNE with 8 scale factors with 1024 shots and 12 collisions.}
    \label{fig:uncorrelated_case_zne}
\end{figure}

\FloatBarrier

\begin{figure}[h]
      \includegraphics[width=0.9\linewidth]{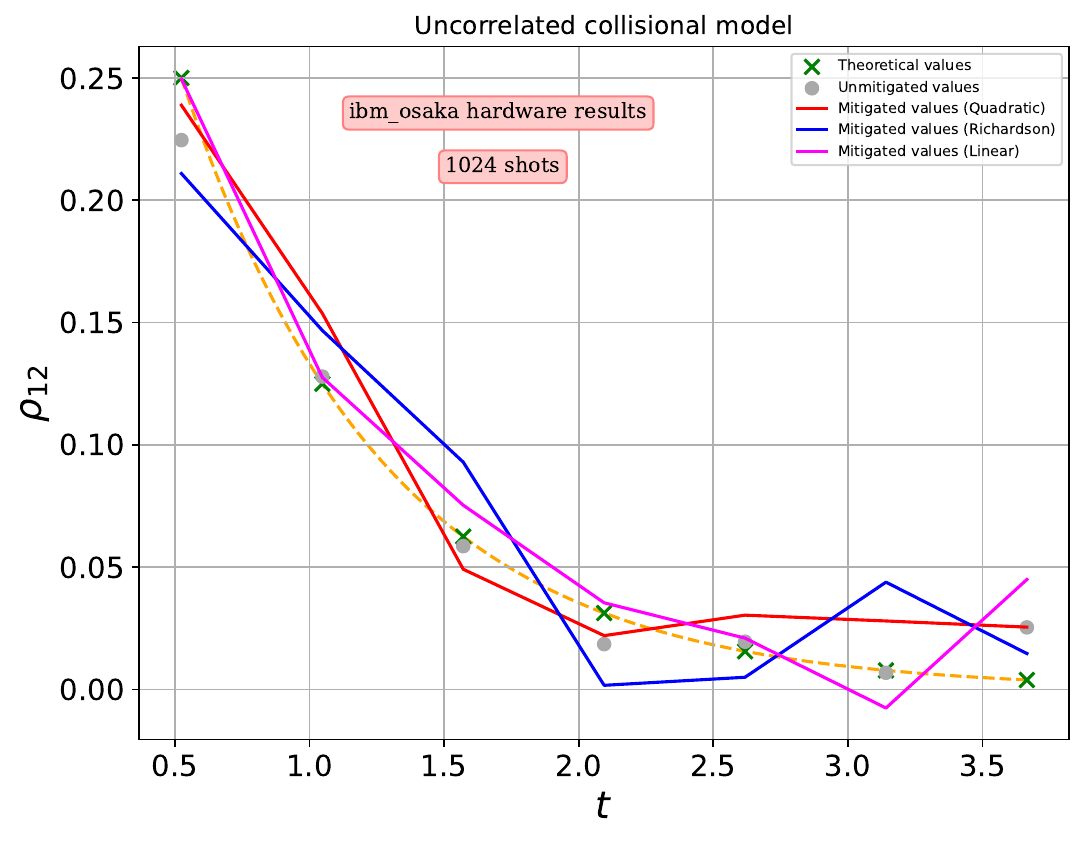}  
      \includegraphics[width=0.9\linewidth]{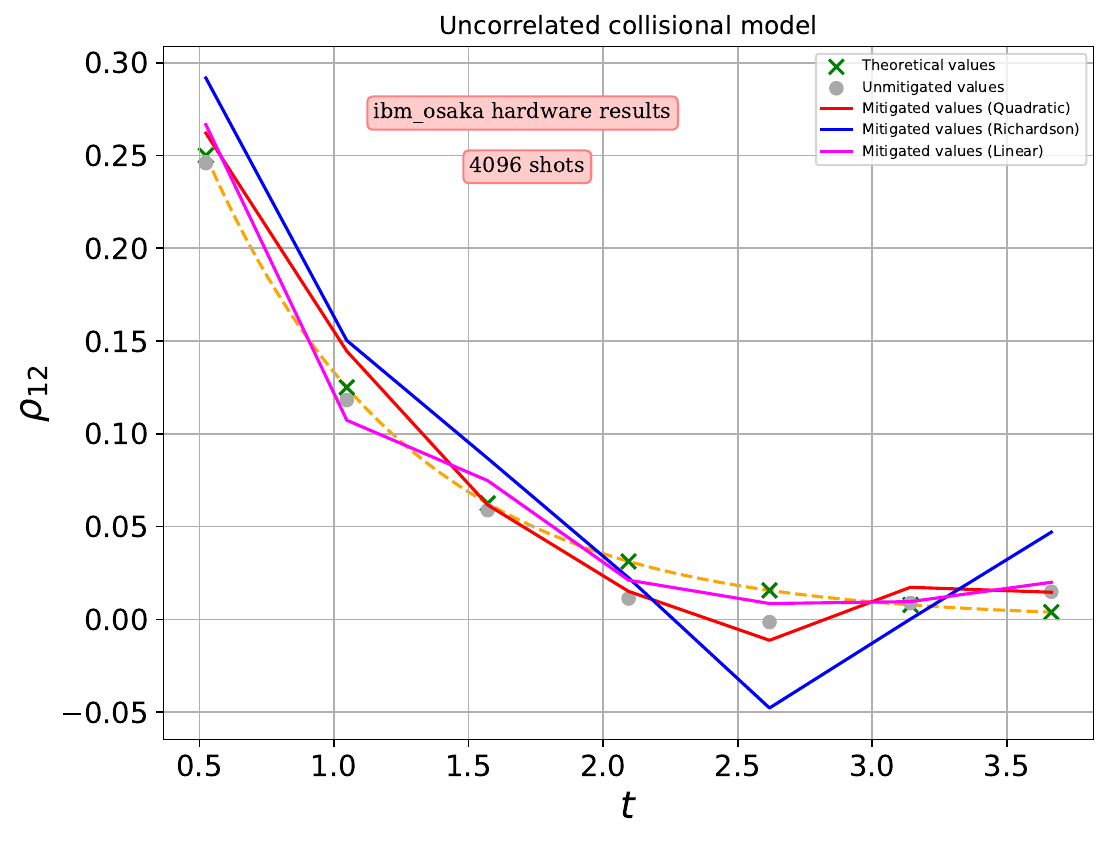} 
  \caption{Results from \texttt{ibmq\_osaka} for the uncorrelated collisional model with ZNE using different extrapolation methods with (Top) 1024 and (Bottom) 4096 shots.The grey points correspond to the experimental, while the dashed lines show the theoretical prediction.}
\label{fig:uncorrelated_case_osaka_results}
\end{figure}

In figure \ref{fig:uncorrelated_case_zne} we can observe the results with noisy simulations using ZNE with a linear extrapolation using 8 scale factors. In this case the results using the mitigation technique doesn't show an improvement compared with the unmitigated one. This results may be because of the number of qubits necessary to simulated each collision in the uncorrelated case.

In figure \ref{fig:uncorrelated_case_osaka_results}, we observe a comparison between different extrapolation methods for scale factors of [1, 3, 5, 7]. The experimental results are from \texttt{ibm\_osaka} with 1024 and 4096 number of shots. The Richardson extrapolations shows the best results. As with the Markovian model simulations, the difference in parameters using ZNE with the noise simulations and the real quantum hardware are because of the complexity of real quantum noise. Also, using REM shows not significant change in the results. 
We can observe an improvement in the error mitigation when increasing the number of shots but still the results are not stable, given the number of qubits needed for each collision.  Comparing with the results in \cite{garcia2020ibm}, we observe an an overall improvement in the values with less shots. Given the number of qubits and the levels of noise of OQC Lucy, this model was not run on their device.


\section{Conclusion}

We have updated and expanded the results of García-Pérez, et al. \cite{garcia2020ibm}, incorporating error mitigation techniques into the experiments and using the latest quantum devices.
The results show that it is not only possible to conduct small experiments on Open Quantum Systems using gate-based NISQ devices, but also the exponential improvements in the performance of these devices and their cloud services by comparing with previous results. The noise simulations, which take into account the gate errors in both the CNOT and one-qubit gates, provide a simple way to study the circuit's behavior in a real QPU before sending it to the cloud service. However, the parameters required for error mitigation may vary depending on the actual noise present in the quantum hardware. Our results indicate a decrease in the error rate caused by noisy qubits when Error Mitigation techniques are employed. In this case, we used the Zero-Noise Extrapolation method for the Markovian Reservoir and collisional models. For the uncorrelated collisional model, the results indicate instability when the number of collisions (qubits) increases, which is consistent with the current NISQ devices.

The experiments reveal that IBM's hardware provides highly accurate results even without error mitigation, whereas OQC hardware produces unstable results due to its noise, which is consistent with previous quantum volume benchmarking of the system \cite{pelofske2022quantum}. We expect that as cloud access to quantum devices expands, we will have greater opportunities to explore larger Open Quantum Systems with different types of hardware.


\section*{Acknowledgments}
This work was supported by qbraid and resulted from the Quantum Open Source Foundation (QOSF) mentorship program.

\bibliography{ref}

\begin{thebibliography}{17}%
\makeatletter
\providecommand \@ifxundefined [1]{%
 \@ifx{#1\undefined}
}%
\providecommand \@ifnum [1]{%
 \ifnum #1\expandafter \@firstoftwo
 \else \expandafter \@secondoftwo
 \fi
}%
\providecommand \@ifx [1]{%
 \ifx #1\expandafter \@firstoftwo
 \else \expandafter \@secondoftwo
 \fi
}%
\providecommand \natexlab [1]{#1}%
\providecommand \enquote  [1]{``#1''}%
\providecommand \bibnamefont  [1]{#1}%
\providecommand \bibfnamefont [1]{#1}%
\providecommand \citenamefont [1]{#1}%
\providecommand \href@noop [0]{\@secondoftwo}%
\providecommand \href [0]{\begingroup \@sanitize@url \@href}%
\providecommand \@href[1]{\@@startlink{#1}\@@href}%
\providecommand \@@href[1]{\endgroup#1\@@endlink}%
\providecommand \@sanitize@url [0]{\catcode `\\12\catcode `\$12\catcode
  `\&12\catcode `\#12\catcode `\^12\catcode `\_12\catcode `\%12\relax}%
\providecommand \@@startlink[1]{}%
\providecommand \@@endlink[0]{}%
\providecommand \url  [0]{\begingroup\@sanitize@url \@url }%
\providecommand \@url [1]{\endgroup\@href {#1}{\urlprefix }}%
\providecommand \urlprefix  [0]{URL }%
\providecommand \Eprint [0]{\href }%
\providecommand \doibase [0]{http://dx.doi.org/}%
\providecommand \selectlanguage [0]{\@gobble}%
\providecommand \bibinfo  [0]{\@secondoftwo}%
\providecommand \bibfield  [0]{\@secondoftwo}%
\providecommand \translation [1]{[#1]}%
\providecommand \BibitemOpen [0]{}%
\providecommand \bibitemStop [0]{}%
\providecommand \bibitemNoStop [0]{.\EOS\space}%
\providecommand \EOS [0]{\spacefactor3000\relax}%
\providecommand \BibitemShut  [1]{\csname bibitem#1\endcsname}%
\let\auto@bib@innerbib\@empty
\bibitem [{\citenamefont {Garc{\'\i}a-P{\'e}rez}\ \emph
  {et~al.}(2020)\citenamefont {Garc{\'\i}a-P{\'e}rez}, \citenamefont {Rossi},\
  and\ \citenamefont {Maniscalco}}]{garcia2020ibm}%
  \BibitemOpen
  \bibfield  {author} {\bibinfo {author} {\bibfnamefont {G.}~\bibnamefont
  {Garc{\'\i}a-P{\'e}rez}}, \bibinfo {author} {\bibfnamefont {M.~A.}\
  \bibnamefont {Rossi}}, \ and\ \bibinfo {author} {\bibfnamefont
  {S.}~\bibnamefont {Maniscalco}},\ }\href@noop {} {\bibfield  {journal}
  {\bibinfo  {journal} {npj Quantum Information}\ }\textbf {\bibinfo {volume}
  {6}},\ \bibinfo {pages} {1} (\bibinfo {year} {2020})}\BibitemShut {NoStop}%
\bibitem [{\citenamefont {Giurgica-Tiron}\ \emph {et~al.}(2020)\citenamefont
  {Giurgica-Tiron}, \citenamefont {Hindy}, \citenamefont {LaRose},
  \citenamefont {Mari},\ and\ \citenamefont {Zeng}}]{giurgica2020digital}%
  \BibitemOpen
  \bibfield  {author} {\bibinfo {author} {\bibfnamefont {T.}~\bibnamefont
  {Giurgica-Tiron}}, \bibinfo {author} {\bibfnamefont {Y.}~\bibnamefont
  {Hindy}}, \bibinfo {author} {\bibfnamefont {R.}~\bibnamefont {LaRose}},
  \bibinfo {author} {\bibfnamefont {A.}~\bibnamefont {Mari}}, \ and\ \bibinfo
  {author} {\bibfnamefont {W.~J.}\ \bibnamefont {Zeng}},\ }in\ \href@noop {}
  {\emph {\bibinfo {booktitle} {2020 IEEE International Conference on Quantum
  Computing and Engineering (QCE)}}}\ (\bibinfo {organization} {IEEE},\
  \bibinfo {year} {2020})\ pp.\ \bibinfo {pages} {306--316}\BibitemShut
  {NoStop}%
\bibitem [{\citenamefont {LaRose}\ \emph {et~al.}(2022)\citenamefont {LaRose},
  \citenamefont {Mari}, \citenamefont {Kaiser}, \citenamefont {Karalekas},
  \citenamefont {Alves}, \citenamefont {Czarnik}, \citenamefont {El~Mandouh},
  \citenamefont {Gordon}, \citenamefont {Hindy}, \citenamefont {Robertson}
  \emph {et~al.}}]{larose2022mitiq}%
  \BibitemOpen
  \bibfield  {author} {\bibinfo {author} {\bibfnamefont {R.}~\bibnamefont
  {LaRose}}, \bibinfo {author} {\bibfnamefont {A.}~\bibnamefont {Mari}},
  \bibinfo {author} {\bibfnamefont {S.}~\bibnamefont {Kaiser}}, \bibinfo
  {author} {\bibfnamefont {P.~J.}\ \bibnamefont {Karalekas}}, \bibinfo {author}
  {\bibfnamefont {A.~A.}\ \bibnamefont {Alves}}, \bibinfo {author}
  {\bibfnamefont {P.}~\bibnamefont {Czarnik}}, \bibinfo {author} {\bibfnamefont
  {M.}~\bibnamefont {El~Mandouh}}, \bibinfo {author} {\bibfnamefont {M.~H.}\
  \bibnamefont {Gordon}}, \bibinfo {author} {\bibfnamefont {Y.}~\bibnamefont
  {Hindy}}, \bibinfo {author} {\bibfnamefont {A.}~\bibnamefont {Robertson}},
  \emph {et~al.},\ }\href@noop {} {\bibfield  {journal} {\bibinfo  {journal}
  {Quantum}\ }\textbf {\bibinfo {volume} {6}},\ \bibinfo {pages} {774}
  (\bibinfo {year} {2022})}\BibitemShut {NoStop}%
\bibitem [{\citenamefont {Hill}\ \emph {et~al.}(2023)\citenamefont {Hill},
  \citenamefont {Young}, \citenamefont {Weis}, \citenamefont {Jun~Liang},
  \citenamefont {Jacobson}, \citenamefont {Gupta}, \citenamefont {Louamri},
  \citenamefont {McIrvin}, \citenamefont {Purohit}, \citenamefont {Necaise},
  \citenamefont {Vara}, \citenamefont {Chakraborty}, \citenamefont {Liu},
  \citenamefont {Coladangelo}, \citenamefont {Kakhandiki}, \citenamefont
  {Makhanov},\ and\ \citenamefont
  {Setia}}]{Hill_qBraid-SDK_Python_toolkit_2023}%
  \BibitemOpen
  \bibfield  {author} {\bibinfo {author} {\bibfnamefont {R.~J.}\ \bibnamefont
  {Hill}}, \bibinfo {author} {\bibfnamefont {R.}~\bibnamefont {Young}},
  \bibinfo {author} {\bibfnamefont {E.}~\bibnamefont {Weis}}, \bibinfo {author}
  {\bibfnamefont {T.}~\bibnamefont {Jun~Liang}}, \bibinfo {author}
  {\bibfnamefont {G.}~\bibnamefont {Jacobson}}, \bibinfo {author}
  {\bibfnamefont {H.}~\bibnamefont {Gupta}}, \bibinfo {author} {\bibfnamefont
  {M.~M.}\ \bibnamefont {Louamri}}, \bibinfo {author} {\bibfnamefont
  {C.}~\bibnamefont {McIrvin}}, \bibinfo {author} {\bibfnamefont
  {S.}~\bibnamefont {Purohit}}, \bibinfo {author} {\bibfnamefont
  {J.}~\bibnamefont {Necaise}}, \bibinfo {author} {\bibfnamefont {E.~A.}\
  \bibnamefont {Vara}}, \bibinfo {author} {\bibfnamefont {P.}~\bibnamefont
  {Chakraborty}}, \bibinfo {author} {\bibfnamefont {J.}~\bibnamefont {Liu}},
  \bibinfo {author} {\bibfnamefont {A.~W.}\ \bibnamefont {Coladangelo}},
  \bibinfo {author} {\bibfnamefont {P.}~\bibnamefont {Kakhandiki}}, \bibinfo
  {author} {\bibfnamefont {H.}~\bibnamefont {Makhanov}}, \ and\ \bibinfo
  {author} {\bibfnamefont {K.}~\bibnamefont {Setia}},\ }\href
  {https://github.com/qBraid/qBraid} {\enquote {\bibinfo {title} {{qBraid-SDK:
  Python toolkit for cross-framework abstraction of quantum programs.}}}\ }
  (\bibinfo {year} {2023})\BibitemShut {NoStop}%
\bibitem [{\citenamefont {Developers}(2023)}]{cirq_developers_2023_10247207}%
  \BibitemOpen
  \bibfield  {author} {\bibinfo {author} {\bibfnamefont {C.}~\bibnamefont
  {Developers}},\ }\href {\doibase 10.5281/zenodo.10247207} {\enquote {\bibinfo
  {title} {Cirq},}\ } (\bibinfo {year} {2023})\BibitemShut {NoStop}%
\bibitem [{\citenamefont {{Qiskit contributors}}(2023)}]{Qiskit}%
  \BibitemOpen
  \bibfield  {author} {\bibinfo {author} {\bibnamefont {{Qiskit
  contributors}}},\ }\href {\doibase 10.5281/zenodo.2573505} {\enquote
  {\bibinfo {title} {Qiskit: An open-source framework for quantum computing},}\
  } (\bibinfo {year} {2023})\BibitemShut {NoStop}%
\bibitem [{\citenamefont {{IBM Quantum}}(2023)}]{IBM_quantum}%
  \BibitemOpen
  \bibfield  {author} {\bibinfo {author} {\bibnamefont {{IBM Quantum}}},\
  }\href {https://quantum.ibm.com/} {\enquote {\bibinfo {title} {{IBM
  Quantum}},}\ } (\bibinfo {year} {2023})\BibitemShut {NoStop}%
\bibitem [{\citenamefont {{Amazon Web Services}}(2020)}]{braket}%
  \BibitemOpen
  \bibfield  {author} {\bibinfo {author} {\bibnamefont {{Amazon Web
  Services}}},\ }\href {https://aws.amazon.com/braket/} {\enquote {\bibinfo
  {title} {{Amazon Braket}},}\ } (\bibinfo {year} {2020})\BibitemShut {NoStop}%
\bibitem [{\citenamefont {{Oxford Quantum Circuits}}(2023)}]{OQC_quantum}%
  \BibitemOpen
  \bibfield  {author} {\bibinfo {author} {\bibnamefont {{Oxford Quantum
  Circuits}}},\ }\href {https://aws.amazon.com/braket/quantum-computers/oqc/}
  {\enquote {\bibinfo {title} {{Oxford Quantum Circuits}},}\ } (\bibinfo {year}
  {2023})\BibitemShut {NoStop}%
\bibitem [{\citenamefont {Barreiro}\ \emph {et~al.}(2011)\citenamefont
  {Barreiro}, \citenamefont {M{\"u}ller}, \citenamefont {Schindler},
  \citenamefont {Nigg}, \citenamefont {Monz}, \citenamefont {Chwalla},
  \citenamefont {Hennrich}, \citenamefont {Roos}, \citenamefont {Zoller},\ and\
  \citenamefont {Blatt}}]{barreiro2011open}%
  \BibitemOpen
  \bibfield  {author} {\bibinfo {author} {\bibfnamefont {J.~T.}\ \bibnamefont
  {Barreiro}}, \bibinfo {author} {\bibfnamefont {M.}~\bibnamefont
  {M{\"u}ller}}, \bibinfo {author} {\bibfnamefont {P.}~\bibnamefont
  {Schindler}}, \bibinfo {author} {\bibfnamefont {D.}~\bibnamefont {Nigg}},
  \bibinfo {author} {\bibfnamefont {T.}~\bibnamefont {Monz}}, \bibinfo {author}
  {\bibfnamefont {M.}~\bibnamefont {Chwalla}}, \bibinfo {author} {\bibfnamefont
  {M.}~\bibnamefont {Hennrich}}, \bibinfo {author} {\bibfnamefont {C.~F.}\
  \bibnamefont {Roos}}, \bibinfo {author} {\bibfnamefont {P.}~\bibnamefont
  {Zoller}}, \ and\ \bibinfo {author} {\bibfnamefont {R.}~\bibnamefont
  {Blatt}},\ }\href@noop {} {\bibfield  {journal} {\bibinfo  {journal}
  {Nature}\ }\textbf {\bibinfo {volume} {470}},\ \bibinfo {pages} {486}
  (\bibinfo {year} {2011})}\BibitemShut {NoStop}%
\bibitem [{\citenamefont {Filippov}\ \emph {et~al.}(2017)\citenamefont
  {Filippov}, \citenamefont {Piilo}, \citenamefont {Maniscalco},\ and\
  \citenamefont {Ziman}}]{filippov2017divisibility}%
  \BibitemOpen
  \bibfield  {author} {\bibinfo {author} {\bibfnamefont {S.~N.}\ \bibnamefont
  {Filippov}}, \bibinfo {author} {\bibfnamefont {J.}~\bibnamefont {Piilo}},
  \bibinfo {author} {\bibfnamefont {S.}~\bibnamefont {Maniscalco}}, \ and\
  \bibinfo {author} {\bibfnamefont {M.}~\bibnamefont {Ziman}},\ }\href@noop {}
  {\bibfield  {journal} {\bibinfo  {journal} {Physical Review A}\ }\textbf
  {\bibinfo {volume} {96}},\ \bibinfo {pages} {032111} (\bibinfo {year}
  {2017})}\BibitemShut {NoStop}%
\bibitem [{\citenamefont {Georgopoulos}\ \emph {et~al.}(2021)\citenamefont
  {Georgopoulos}, \citenamefont {Emary},\ and\ \citenamefont
  {Zuliani}}]{georgopoulos2021modeling}%
  \BibitemOpen
  \bibfield  {author} {\bibinfo {author} {\bibfnamefont {K.}~\bibnamefont
  {Georgopoulos}}, \bibinfo {author} {\bibfnamefont {C.}~\bibnamefont {Emary}},
  \ and\ \bibinfo {author} {\bibfnamefont {P.}~\bibnamefont {Zuliani}},\
  }\href@noop {} {\bibfield  {journal} {\bibinfo  {journal} {Physical Review
  A}\ }\textbf {\bibinfo {volume} {104}},\ \bibinfo {pages} {062432} (\bibinfo
  {year} {2021})}\BibitemShut {NoStop}%
\bibitem [{\citenamefont {Temme}\ \emph {et~al.}(2017)\citenamefont {Temme},
  \citenamefont {Bravyi},\ and\ \citenamefont {Gambetta}}]{temme2017error}%
  \BibitemOpen
  \bibfield  {author} {\bibinfo {author} {\bibfnamefont {K.}~\bibnamefont
  {Temme}}, \bibinfo {author} {\bibfnamefont {S.}~\bibnamefont {Bravyi}}, \
  and\ \bibinfo {author} {\bibfnamefont {J.~M.}\ \bibnamefont {Gambetta}},\
  }\href@noop {} {\bibfield  {journal} {\bibinfo  {journal} {Physical review
  letters}\ }\textbf {\bibinfo {volume} {119}},\ \bibinfo {pages} {180509}
  (\bibinfo {year} {2017})}\BibitemShut {NoStop}%
\bibitem [{\citenamefont {Bravyi}\ \emph {et~al.}(2021)\citenamefont {Bravyi},
  \citenamefont {Sheldon}, \citenamefont {Kandala}, \citenamefont {Mckay},\
  and\ \citenamefont {Gambetta}}]{bravyi2021mitigating}%
  \BibitemOpen
  \bibfield  {author} {\bibinfo {author} {\bibfnamefont {S.}~\bibnamefont
  {Bravyi}}, \bibinfo {author} {\bibfnamefont {S.}~\bibnamefont {Sheldon}},
  \bibinfo {author} {\bibfnamefont {A.}~\bibnamefont {Kandala}}, \bibinfo
  {author} {\bibfnamefont {D.~C.}\ \bibnamefont {Mckay}}, \ and\ \bibinfo
  {author} {\bibfnamefont {J.~M.}\ \bibnamefont {Gambetta}},\ }\href@noop {}
  {\bibfield  {journal} {\bibinfo  {journal} {Physical Review A}\ }\textbf
  {\bibinfo {volume} {103}},\ \bibinfo {pages} {042605} (\bibinfo {year}
  {2021})}\BibitemShut {NoStop}%
\bibitem [{Note1()}]{Note1}%
  \BibitemOpen
  \bibinfo {note} {\protect \url
  {https://github.com/JessicaJohnBritto/QOSF_Cohort6}}\BibitemShut {NoStop}%
\bibitem [{\citenamefont {Abu-Nada}\ and\ \citenamefont
  {Salhab}(2023)}]{abu2023preparing}%
  \BibitemOpen
  \bibfield  {author} {\bibinfo {author} {\bibfnamefont {A.~A.}\ \bibnamefont
  {Abu-Nada}}\ and\ \bibinfo {author} {\bibfnamefont {M.~A.}\ \bibnamefont
  {Salhab}},\ }\href@noop {} {\bibfield  {journal} {\bibinfo  {journal}
  {International Journal of Quantum Information}\ ,\ \bibinfo {pages}
  {2340008}} (\bibinfo {year} {2023})}\BibitemShut {NoStop}%
\bibitem [{\citenamefont {Pelofske}\ \emph {et~al.}(2022)\citenamefont
  {Pelofske}, \citenamefont {B{\"a}rtschi},\ and\ \citenamefont
  {Eidenbenz}}]{pelofske2022quantum}%
  \BibitemOpen
  \bibfield  {author} {\bibinfo {author} {\bibfnamefont {E.}~\bibnamefont
  {Pelofske}}, \bibinfo {author} {\bibfnamefont {A.}~\bibnamefont
  {B{\"a}rtschi}}, \ and\ \bibinfo {author} {\bibfnamefont {S.}~\bibnamefont
  {Eidenbenz}},\ }\href@noop {} {\bibfield  {journal} {\bibinfo  {journal}
  {IEEE Transactions on Quantum Engineering}\ }\textbf {\bibinfo {volume}
  {3}},\ \bibinfo {pages} {1} (\bibinfo {year} {2022})}\BibitemShut {NoStop}%
\end{thebibliography}%

\section{Appendices} \label{Sect:MRE_kraus} 

\section{Architecture of Quantum Hardware}

\begin{figure}[h]
      \includegraphics[width=0.8\linewidth]{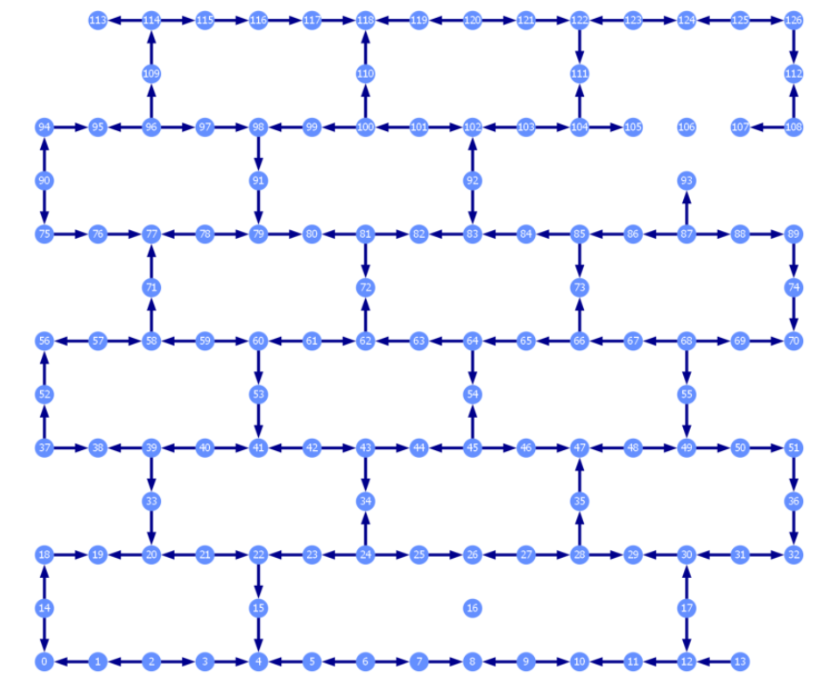} 
  \caption{ Architecture of \texttt{ibm\_osaka}}
  \label{fig:Architure_QH_Used_ibm_osaka}
\end{figure}

\begin{figure}[h]
      \includegraphics[width=0.8\linewidth]{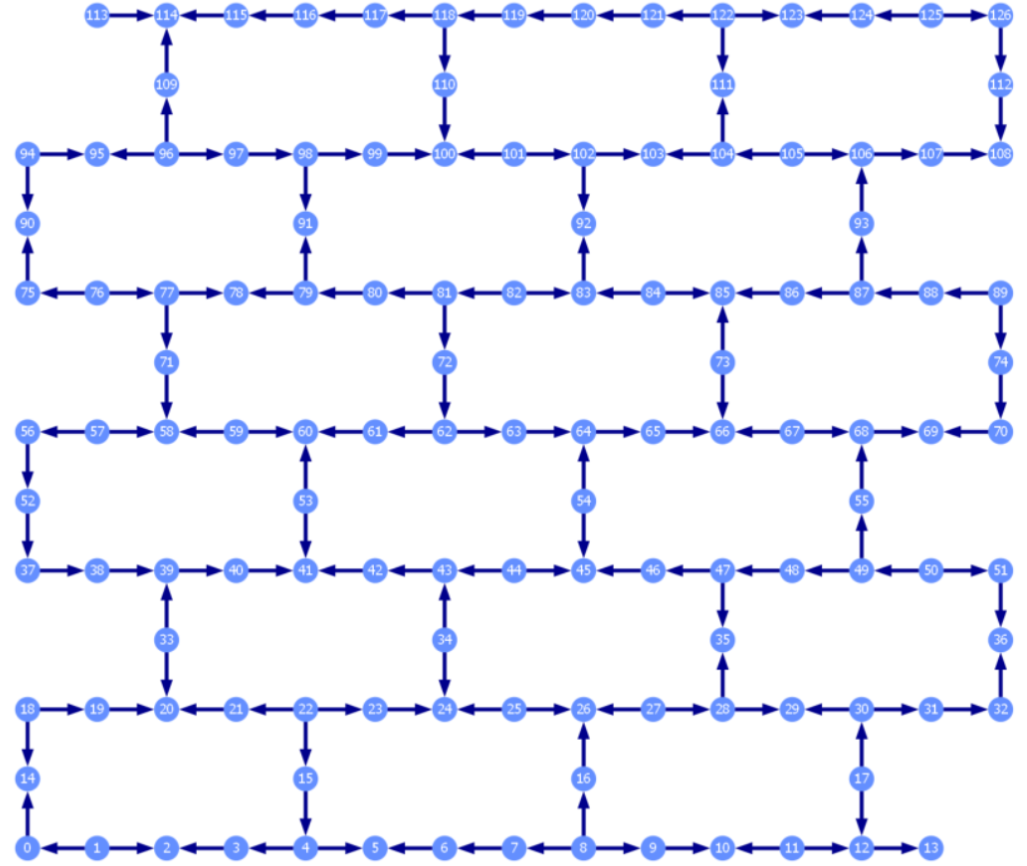} 
  \caption{ Architecture of \texttt{ibm\_kyoto}}
  \label{fig:Architure_QH_Used_ibm_tokyo}
\end{figure}

\begin{figure}[h]
      \includegraphics[width=0.45\linewidth]{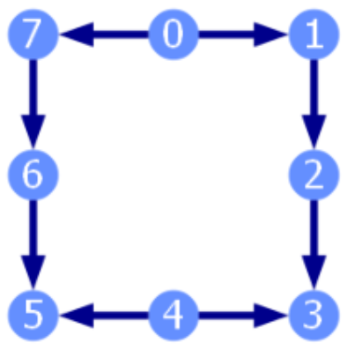}  
  \caption{Architecture of  OQC Lucy}
  \label{fig:Architure_QH_Used_lucy}
\end{figure}

In figure \ref{fig:Architure_QH_Used_ibm_osaka} and \ref{fig:Architure_QH_Used_ibm_tokyo}, we can see the topology of the IBM hardware used for the experiments. For OQC Lucy, we can see in figure \ref{fig:Architure_QH_Used_lucy} the topology of the device. In the case of Lucy, the two qubit connections are different than with the IBM devices. The calibration data for the IBM hadware is shown in table \ref{table:specs_osaka} and \ref{table:specs_kyoto}, for OQC Lucy in table \ref{table:lucy_specs} 

\begin{table}[h!]
\centering
\begin{tabular}{ |c|c| } 
 \hline
 Property & Median\\ 
 \hline
 T1 ($\mu s$) & 270.36 \\ 
\hline
 T2 ($\mu s$) & 170.43 \\ 
 \hline
 SX error & 1.800e-2\\ 
 \hline
 ECR error & 8.257e-3 \\ 
 \hline
 Readout error & 1.8e-2\\
 \hline
\end{tabular}
  \caption{Calibration data for  \texttt{ibm\_osaka} from IBM quantum \cite{IBM_quantum} with gate basis [ECR, ID, RZ, SX, X]. }
\label{table:specs_osaka}
\end{table}

\begin{table}[h!]
\centering
\begin{tabular}{ |c|c| } 
 \hline
 Property & Median\\ 
 \hline
 T1 ($\mu s$) & 223.49 \\ 
\hline
 T2 ($\mu s$) & 118.92 \\ 
 \hline
 SX error & 2.484e-2\\ 
 \hline
 ECR error & 8.12e-3 \\ 
 \hline
 Readout error & 1.53e-2\\
 \hline
\end{tabular}
  \caption{Calibration data for  \texttt{ibm\_kyoto} from IBM quantum \cite{IBM_quantum}  with gate basis [ECR, ID, RZ, SX, X]. }
\label{table:specs_kyoto}
\end{table}

\FloatBarrier

\begin{table}[h!]
\centering
\begin{tabular}{ |c|c|c| } 
 \hline
 Property & Average  & Median\\ 
 \hline
 T1 ($\mu s$) & 36.460 & 34.469\\ 
\hline
 T2 ($\mu s$) & 33.401 & 37.972 \\ 
 \hline
 Fidelity (RB) (\%) & 99.886 & 99.912\\ 
 \hline
 Readout Fidelity (\%) & 88.675 & 89.250 \\ 
 \hline
 ECR gate fidelity (\%) & 94.672 & 95.589\\
 \hline
\end{tabular}
  \caption{Calibration data for OQC Lucy from AWS services \cite{braket} with gate basis [ect, i, rz, v, x]. }
\label{table:lucy_specs}
\end{table}

\end{document}